\documentclass[reprint,aps,amsmath,amssymb,superscriptaddress,twocolumn,notitlepage,nofootinbib,longbibliography]{revtex4-2}

\usepackage[final]{graphicx}
\usepackage{times,bbm,amsmath,amssymb,amsthm}
\usepackage{epsfig,color}
\usepackage[dvipsnames]{xcolor}
\usepackage[colorlinks=true,citecolor=blue,linkcolor=blue]{hyperref}
\hypersetup{
    allcolors  = {blue},
}

\usepackage{cleveref}

\usepackage{float}
\usepackage[caption = false]{subfig}
\usepackage{verbatim}
\usepackage[greek,english]{babel}
\usepackage{thumbpdf,enumerate}
\usepackage{booktabs}
\usepackage{sidecap}
\usepackage[scaled=.8]{couriers}    
\usepackage{pstricks}
\usepackage{multirow}
\usepackage{placeins}
\usepackage{relsize}  
\usepackage{pst-grad,bm}
\usepackage{epigraph}
\usepackage{gensymb}
\usepackage{longtable}
\usepackage{booktabs}
\usepackage{gensymb}

\usepackage{soul}
\usepackage{ulem} 
\usepackage{acronym}
\usepackage{physics}
\usepackage{tikz}

\usepackage{todonotes}

\graphicspath{{Images/}}

\newcommand{\EV}[1]{\langle{#1}\rangle}

\begin{document}

\title{Observation of Lie algebraic invariants in Quantum Linear Optics}

\author{Giovanni Rodari}
\affiliation{Dipartimento di Fisica, Sapienza Universit\`{a} di Roma, Piazzale Aldo Moro 5, I-00185 Roma, Italy}

\author{Tommaso Francalanci}
\affiliation{Dipartimento di Fisica, Sapienza Universit\`{a} di Roma, Piazzale Aldo Moro 5, I-00185 Roma, Italy}

\author{Eugenio Caruccio}
\affiliation{Dipartimento di Fisica, Sapienza Universit\`{a} di Roma, Piazzale Aldo Moro 5, I-00185 Roma, Italy}

\author{Francesco Hoch}
\affiliation{Dipartimento di Fisica, Sapienza Universit\`{a} di Roma, Piazzale Aldo Moro 5, I-00185 Roma, Italy}

\author{Gonzalo Carvacho}
\affiliation{Dipartimento di Fisica, Sapienza Universit\`{a} di Roma, Piazzale Aldo Moro 5, I-00185 Roma, Italy}

\author{Taira Giordani}
\affiliation{Dipartimento di Fisica, Sapienza Universit\`{a} di Roma, Piazzale Aldo Moro 5, I-00185 Roma, Italy}

\author{Nicol\`o Spagnolo}
\affiliation{Dipartimento di Fisica, Sapienza Universit\`{a} di Roma, Piazzale Aldo Moro 5, I-00185 Roma, Italy}

\author{Riccardo Albiero}
\affiliation{Istituto di Fotonica e Nanotecnologie, Consiglio Nazionale delle Ricerche (IFN-CNR), Piazza Leonardo da Vinci, 32, I-20133 Milano, Italy}

\author{Niki Di Giano}
\affiliation{Istituto di Fotonica e Nanotecnologie, Consiglio Nazionale delle Ricerche (IFN-CNR), 
Piazza Leonardo da Vinci, 32, I-20133 Milano, Italy}
\affiliation{Dipartimento di Fisica - Politecnico di Milano, Piazza Leonardo da Vinci 32, 20133 Milano, Italy}

\author{Francesco Ceccarelli}
\affiliation{Istituto di Fotonica e Nanotecnologie, Consiglio Nazionale delle Ricerche (IFN-CNR), Piazza Leonardo da Vinci, 32, I-20133 Milano, Italy}

\author{Giacomo Corrielli}
\affiliation{Istituto di Fotonica e Nanotecnologie, Consiglio Nazionale delle Ricerche (IFN-CNR), Piazza Leonardo da Vinci, 32, I-20133 Milano, Italy}

\author{Andrea Crespi}
\affiliation{Istituto di Fotonica e Nanotecnologie, Consiglio Nazionale delle Ricerche (IFN-CNR), Piazza Leonardo da Vinci, 32, I-20133 Milano, Italy}
\affiliation{Dipartimento di Fisica - Politecnico di Milano, Piazza Leonardo da Vinci 32, 20133 Milano, Italy}

\author{Roberto Osellame}
\affiliation{Istituto di Fotonica e Nanotecnologie, Consiglio Nazionale delle Ricerche (IFN-CNR), Piazza Leonardo da Vinci, 32, I-20133 Milano, Italy}

\author{Ulysse Chabaud}
\affiliation{DIENS, \'Ecole Normale Sup\'erieure, PSL University, CNRS, INRIA, 45 rue d’Ulm, Paris, 75005, France}

\author{Fabio Sciarrino}
\email{fabio.sciarrino@uniroma1.it}
\affiliation{Dipartimento di Fisica, Sapienza Universit\`{a} di Roma, Piazzale Aldo Moro 5, I-00185 Roma, Italy}

\begin{abstract}
Over the past few years, various methods have been developed to engineeer and to exploit the dynamics of photonic quantum states as they evolve through linear optical networks.
Recent theoretical works have shown that the underlying Lie algebraic structure plays a crucial role in the description of linear optical Hamiltonians, as such formalism identifies intrinsic symmetries within photonic systems subject to linear optical dynamics.
Here, we experimentally investigate the role of Lie algebra applied to the context of Boson sampling, a pivotal model to the current understanding of computational complexity regimes in photonic quantum information.
Performing experiments of increasing complexity, realized within a fully-reconfigurable photonic circuit, we show that sampling experiments do indeed fulfill the constraints implied by a Lie algebraic structure. In addition, we provide a comprehensive picture about how the concept of Lie algebraic invariant can be interpreted from the point of view of n-th order correlation functions in quantum optics.
Our work shows how Lie algebraic invariants can be used as a benchmark tool for the correctness of an underlying linear optical dynamics and to verify the reliability of Boson Sampling experiments. This opens new avenues for the use of algebraic-inspired methods as verification tools for photon-based quantum computing protocols.
\end{abstract}

\maketitle

\section{Introduction}

Harnessing linear optical systems has been a cornerstone of photonic-based quantum information processing, finding several applications \cite{Kok2007,Motes2015,Ewert2016,Slussarenko2019,Flamini2018}. In particular, Boson Sampling and its variants \cite{Aaronson2011,brod2019photonic,lund2014boson,hamilton2017gaussian,kruse2019detailed,chabaud2021quantum,spagnolo2023non} have been proposed as a model in which multi-photon states, linear optical evolution, and subsequent photon detection can be employed to reach the quantum advantage regime \cite{aaronson2016complexity,preskill2018quantum} with current photonic technologies \cite{heindel2023quantum,giordani2023integrated,wang2020integrated}. Research efforts in this paradigm have been stimulated by its suitability to observe the enhanced computational power of quantum devices for solving a classically-hard problem \cite{Aaronson2011}, and by the development of photon-based universal quantum computation schemes \cite{knill2001scheme,bartolucci2023fusion}. As demonstrated in increasingly complex schemes \cite{broome2013photonic, spring2013boson, tillmann2013experimental, crespi2013integrated, loredo2017boson, he2017time, wang2017experimental, zhong201812,wang2019boson,zhong2020quantum,madsen2022quantum,deng2023gaussian,hoch2025}, Boson Sampling photonic platforms provide a suitable testbed to investigate novel approaches to analyze quantum computational complexity. In this context, several benchmarking methods have been proposed to quantitatively assess the correct functioning of such quantum devices \cite{rodari2024semi,vandermeer2021witness,Walschaers2016statistical,Giordani2018experimental,aaronson2014far,Tichy2014stringent,Crespi2016suppr,Spagnolo2014vali,Pont2022quantifying,Agresti2019pattern}. 

Recently, concepts stemming from Lie algebras theory \cite{invariants,parellada2023no,garcia2019multiple} have been proposed as tools for the analysis and design of photonics-based quantum information protocols reliant on linear optics. 
The mathematical toolbox of quantum evolution operators describing linear optical networks and scattering matrices is related to Lie algebras by isomorphisms, providing an elegant representation of these optical transformations.
This parallel offers new insights into the dynamics of quantum linear optical evolutions without the need of devising particularly complex measurement setups, i.e., many-mode interactions. Specifically, one can identify Lie algebraic invariants---measurable quantities that are conserved under linear optical transformations---which characterize photonic evolution \cite{invariants}. Due to considerations connected to the theory of Lie algebra, the value of a suitably constructed function of expectation values measured at the output of an optical circuit is conserved whenever the underlying dynamic is linear. Operationally, the lowest order invariants can be computed by measuring the mean photon numbers on a given set of outputs of a linear optical network, or their combination after further manipulation in order to implement two-mode operators.
These objects are instrumental in defining particular constraints and symmetries that do characterize the behavior of linear-based photonic systems, i.e., the framework of Boson Sampling. In particular, this provides a new perspective on the constraints imposed by linear optics on the accessible output states. This is particularly relevant in the context of Boson Sampling, where characterizing the set of reachable distributions helps in assessing its potential for quantum computing applications, such as quantum machine learning in its adaptive variant \cite{chabaud2021quantum, hoch2025}.

In this article, we experimentally investigate the role of invariants of the Lie algebra of linear-optics Hamiltonians within the context of Boson Sampling.
We explore how these invariants can be measured and verified using a state-of-the-art photonic architecture. We validate the feasibility of measuring such quantities in photonic experiments of increasing complexity, employing up to four-photon states obtained via a demultiplexed quantum dot source \cite{somaschi2016near,senellart2017high,heindel2023quantum} and injected in an eight-mode, fully reconfigurable photonic circuit \cite{corrielli2021femtosecond,pentangelo2022universal,pentangelo2024high,clements2016optimal}. 
We are able to accurately program the interferometer to implement a controlled linear optical evolution followed by a set of photon number measurements, required to experimentally estimate a given Lie algebraic quantity. In addition, we show how the operators and invariant quantities of the Lie algebra can be connected to the concepts of quantum correlations and coherence properties of quantum states of light.
Overall, by exploiting Lie algebraic tools within the context of photon-based quantum information processing, we provide a novel insight into the theoretical understanding and experimental validation of linear optical quantum systems.

\section{Background: the role of Lie Algebraic invariants in linear optics}

The dynamical behaviour of photonic quantum states undergoing a general linear optical evolution was characterized, from the point of view of quantum state preparation, in a recent work by Parellada et al. \cite{invariants}. 
By exploiting the connection between the unitary evolution generated by a passive linear optical interferometer and the mathematical structure of Lie algebras \cite{garcia2019multiple}, several invariant quantities are derived for passive linear optical systems. 
Such invariants, all stemming from the projection of the density operator describing the quantum states onto the Lie algebra of passive linear optical Hamiltonians, are conserved when a quantum state evolves through a passive linear optical interferometer. 

Conserved quantities arise when one considers the expectation values over a suitable operator basis describing the algebraic space of linear optical Hamiltonian operators. In particular, the following basis of Hermitian operators, namely physical observables, can be adopted for a system characterized by $m$ optical modes as in Fig.~\ref{fig:liescheme}-(a):
\begin{equation}\label{eq:observables}
\begin{cases}
{O}_j^z = n_j = a^\dagger_j a_j & \text{for } j = 1, \dots, m, \\
{O}_{jk}^x = \frac{1}{\sqrt{2}} 
\left( 
a^\dagger_j a_k + a^\dagger_k a_j 
\right) 
& \text{for } 1 \leq j < k \leq m, \\
{O}_{jk}^y = \frac{i}{\sqrt{2}} 
\left( 
a^\dagger_j a_k - a^\dagger_k a_j 
\right) 
& \text{for } 1 \leq j < k \leq m.    
\end{cases}
\end{equation}
The expectation values of the observables in Eq.~\eqref{eq:observables} can be experimentally measured, with multi-mode interferometers, in the following way: (i) for the $O_j^z$ operators, by probing the populations of the output density matrix expressed in the Fock basis representation; (ii) for the operators $O_{jk}^{x,y}$, by measuring the mean photon number difference $\expval{n_j} - \expval{n_k}$ between modes $j$ and $k$ \cite{campos_beamsplitter} after letting two modes $(j,k)$ interfere on a balanced beam-splitter, with an additional phase term set to $0$ ($\frac{\pi}{2}$) for $O_{jk}^{x}$ ($O_{jk}^{y}$). The schemes to measure such operators for two spatial modes are shown in Fig.~\ref{fig:liescheme}-(b). The construction can be generalized to a higher number of modes, as shown for three (four) spatial modes in Fig.~\ref{fig:liescheme}-(c) [Fig.~\ref{fig:liescheme}-(d)], where additional mode swapping is required to correctly let modes $(j, k)$ interfere.
Note that, at variance from a standard Boson Sampling experiment, the measurement of the observables defined above also requires an additional measurement stage after the linear optical evolution, as depicted in Fig.~\ref{fig:liescheme}.
Moreover, the adopted detectors should have the capability to resolve the number of photons on each detected mode in order to compute the required mean photon number values $\expval{n_k}$.

\subsection{Lie algebraic invariants of interest}

\begin{figure*}[ht!]
    \centering
    \includegraphics[width=0.95\textwidth]{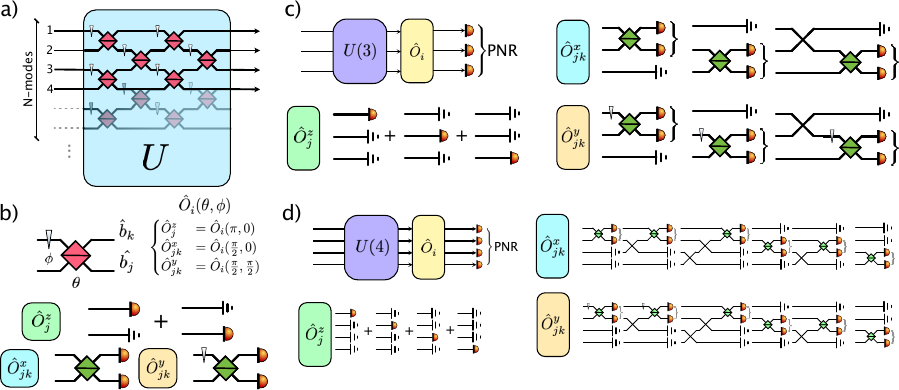}
    \caption{\textbf{Linear optical schemes for measuring Lie algebraic invariants:} a) A $m$-mode unitary transformation constructed only via linear optical elements, variable beam splitters (red squares) and phase shifters (light blue triangles) is provided together with an arbitrary photon input state $\rho$, with $\rho = \ket{\bm{s}}\bra{\bm{s}}$. In our experiment, $\ket{\bm{s}}$ denotes a multi-mode $n$-photon Fock state. The state evolves throughout the interferometer, leading to an output state $\rho'$. b) In order to measure the invariant $I(\rho')$, as defined in Eq. \eqref{eq:invariant}, on the state $\rho'$ at the output of a linear optical dynamic, every possible pair of output modes is let interfere through optical setups involving at most two modes, as depicted in the panel, through which the expectation values of the operators $\{{O}_{j}^z, {O}_{jk}^x, {O}_{jk}^y\}$ on state $\rho'$ can be measured. Here, the green square denotes a variable beam-splitter set to a 50:50 splitting ratio.
    c) The unitary transformation $U(3)$ is followed by mode-swapping set up and the circuits shown in panel (b) in order to measure the expectation values of a given operator, which requires photon number resolving detection. In this a way, the experimental value of the invariant $I(\rho')$ in the $m=3$ scenario can be reconstructed.
    d) Similarly, the experimental value of the invariant $I(\rho)$ in the $m = 4$ scenario can be reconstructed by implementing jointly a linear optical transformation represented by a unitary matrix $U(4)$ followed by the measurements setups required to probe the aforementioned expectation values. Also in this case photon-number resolving detection is operated at the output.}
    \label{fig:liescheme}
\end{figure*}

The first Lie algebraic invariant of interest is obtained as the sum of the squared expectation values of the observables defined in Eq.~\eqref{eq:observables}:  
\begin{equation}
    I(\rho) = \sum_{i=1}^{m^2} \Tr(O_i \rho)^2.
    \label{eq:invariant}
\end{equation}  
This scalar quantity represents the purity of the state projected onto the chosen basis and remains unchanged under linear optical evolution. Considering as input a fixed generic Fock state $\ket{\bm{s}}=\ket{n_1,..., n_m}$ comprising $n = \sum_{j=1}^m n_j$ photons traversing $m$ optical modes, it can be shown that 
\begin{equation}
    I(\dyad{\bm{s}}) = \sum_{j=1}^{m} n_j^2.
    \label{eq:irhofock}
\end{equation}
Notably, for Fock basis pure states comprising $n$ photons distributed in such a way to exhibit at most one photon per mode - i.e. the prototypical input state employed in several recent demonstrations of photon based quantum information protocols - Eq. \eqref{eq:irhofock} reduces to $I(\dyad{\bm{s}}) = n$.
Since the expectation values $\Tr(O_i \rho)$ are experimentally measurable with the apparatus shown in Fig.~\ref{fig:liescheme}, one can directly test whether $I(\rho)$ is conserved for a given input state after optical evolution. If this quantity is not conserved one should conclude that the evolution is non-linear. Moreover, if two states have different values of $I(\rho)$, they cannot be connected by a passive linear optical transformation.

A second Lie algebraic invariant can be obtained considering the infinite-dimensional density operator defined as:
\begin{equation}
    \rho_T = \sum_{i=1}^{m^2} \Tr(O_i \rho)O_i,
    \label{eq:dm_invariant}
\end{equation}
The operator $\rho_T$ represents the projection of the density matrix onto the subalgebra of linear optical Hamiltonians. The eigenvalue spectra of $\rho_T$ is indeed invariant under linear optical evolution. The invariance of its spectrum follows from the fact that linear optical evolution keeps it within the subalgebra, as shown in \cite{invariants}. Note that the spectrum of $\rho_T$ has infinite cardinality and, since $\rho_T$ preserves the total number of photons, it is block-diagonal. In what follows, we will refer to the eigenvalues of each block as $\lambda_n$, with $n$ being the conserved number of photons of each block (see Supplementary Material for more details). This result provides additional constraints beyond $I(\rho)$, as it gives more information than a single scalar value.
The previous physical invariants provide valuable tools for analyzing the possibilities and limitations of state preparation in quantum linear optics. Indeed, they help to constrain the search for the linear unitary evolution preparing a given output photon state, which might be impossible to obtain due to the non-conservation of the associated invariant with respect to a fixed input resource. 

\begin{figure}[ht!]
    \centering
    \includegraphics[width=0.85\linewidth]{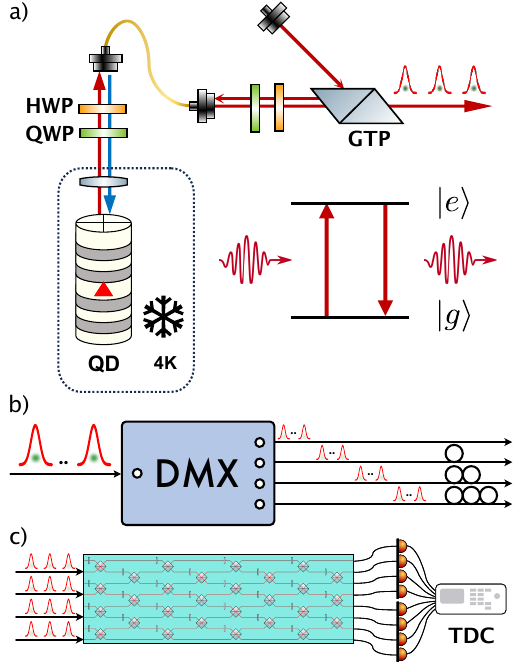}
    \caption{\textbf{The employed state-of-the-art hybrid photonic architecture:} a) A Quantum-Dot (QD) single-photon source is resonantly excited in a cross-polarized configuration where separation of the residual pump laser is achieved by means of a Glan-Thompson polarizer (GTP), Half-wave plates (HWP) and Quarter-wave plates (QWP), to compensate from the polarization rotations induced by the propagation in the optical fiber and control the laser pump polarization state at the interface with the QD.
    b) A time-to-spatial demultiplexer actively splits the input single-photon stream into a fixed set of output modes, where a multi-photon resource state with up to $n=4$ photons is obtained.
    c) The time-to-space demultiplexed source is interfaced with an eight-mode, fully reconfigurable photonic integrated circuit. Here, the circuit is programmed in such a way to implement a linear optical dynamic followed by the optical setup required to measure the Lie operators, as shown in Fig. \ref{fig:liescheme}. At the output of the circuit, photon detection is carried out via superconducting nanowire detectors, connected a time-to-digital converter (TDC) which records $n$-fold coincidence events.}
    \label{fig:exp_scheme}
\end{figure}

\subsection{Connection with the concept of quantum optical coherency matrix}

An interesting insight into the physical meaning of the observables in Eq.~\eqref{eq:observables} can be gained by recalling the concept of quantum optical coherency matrix \cite{fabre2020modes, glauber1963quantum}, an $m \times m$ square matrix with entries:
\begin{equation}
(\mathbf{\Gamma}^{(1)})_{jk} = \EV{ a^\dagger_j  a_k}.
\end{equation}
Such entries are in fact proportional to the first-order quantum correlation functions for the multi-mode field $G^{(1)}_{jk} = \EV{ E^{-}_j  E^{+}_k} = \mathcal{E}^2 \EV{ a^\dagger_j  a_k}$.
The quantum coherency matrix is Hermitian, and may be written more compactly as:
\begin{equation}
\mathbf{\Gamma}^{(1)} = \EV{\vec a^\dagger \vec a^T},
\label{eq:gamma1matrix}
\end{equation}
where $\vec a$ is a column vector containing the annihilation operators of the $m$ modes.
The entries of the coherency matrix are related to the expectation values of the observables \eqref{eq:observables} by the following linear transformations:
\begin{equation}
\begin{aligned}
(\mathbf{\Gamma}^{(1)})_{jk} = (\mathbf{\Gamma}^{(1)})_{kj}^*  &=\frac{1}{\sqrt{2}}\left(\EV{O^x_{jk}} - i \EV{O^y_{jk}}\right) \\
(\mathbf{\Gamma}^{(1)})_{jj} &= \EV{O^z_{j}}
\end{aligned} \label{eq:ojTOg}
\end{equation}
Thus, the experimental estimation of the expectation values $\EV{O_i}$ coincides with the experimental characterization of the quantum coherency matrix of the multi-mode field. It can be further noted that the information contained in the operator $\rho_T$ in Eq.~\eqref{eq:dm_invariant} is the same contained in the coherency matrix, as we can obtain the coefficients $\EV{O_i} = \Tr(O_i \rho)$ from the entries of the coherency matrix and vice versa, using Eq.~\eqref{eq:ojTOg}. 

The effect of a linear optical transformation on the output coherence matrix is equivalent to a basis change in the representation of $\Gamma^{(1)}$ \cite{fabre2020modes}. This follows by direct application of the unitary operator that represents a given linear optical transformation to the bosonic operators in the definition of the coherency matrix as in Eqs. \eqref{eq:gamma1matrix}-\eqref{eq:ojTOg}. In particular, let $U$ be the unitary matrix describing the linear-optics transformation on the set of creation vectors of the optical modes, so that the equalities $\vec b^\dagger = U \vec a^\dagger$ and $\vec b = U^* \vec a$ hold.
Then, a linear optical evolution acts on the coherency matrix as a basis change, according to:
\begin{equation}
\mathbf{\Gamma}^{(1)}_b = \EV{\vec b^\dagger \vec b^T} = U \EV{  \vec a^\dagger \vec a^T } U^\dagger.
\end{equation}

The relation between the expectation values $\EV{O_i}$ and the entries of the coherency matrix in Eq.~(\ref{eq:ojTOg}) allows us to give interesting interpretations of the algebraic invariants discussed above.
Noting that $\EV{O^x_{jk}} = \sqrt{2} \mathrm{Re}\lbrace\EV{a_j^\dagger a_k} \rbrace$ and $\EV{O^y_{jk}} = - \sqrt{2} \mathrm{Im}\lbrace\EV{a_j^\dagger a_k} \rbrace$, one can derive easily that the quantity $I(\rho)$ in Eq.~\eqref{eq:invariant} corresponds to:
\begin{equation}
I(\rho) = \sum_{j,k} \left|\EV{ a_j^\dagger  a_k}\right|^2\! =  \sum_{j,k} \left|(\mathbf{\Gamma}^{(1)})_{jk}\right|^2\! = \mathrm{Tr}\left[ (\mathbf{\Gamma}^{(1)})^\dagger \mathbf{\Gamma}^{(1)} \right]\!.\label{eq:invariantLie1}
\end{equation}
Namely, the invariant $I(\rho)$ corresponds to the Frobenius norm of the coherency matrix. Then, the fact that $I(\rho)$ is an invariant for linear-optics transformations directly corresponds to the conservation of the norm of the coherency matrix $\mathbf{\Gamma}^{(1)}$ upon a change of basis.

Furthermore, as the coherency matrix is Hermitian, there is always an eigenvector basis that diagonalizes it, which is the basis of the \emph{principal modes} \cite{fabre2020modes}. The modes as defined in this basis do not show classical interference with each other, because the non-diagonal elements corresponding to the correlations of the operators $G^{(1)}_{jk}$ with $j \neq k$ vanish in this basis. The eigenvalues of the coherency matrix, namely the diagonal entries when the matrix is diagonalized, are the photon populations $\EV{N_i}$ of such principal modes. The value of $I(\rho)$, interpreted as the squared norm of the matrix, equals the sum of the squares of the eigenvalues, i.e. $I(\rho) = \sum_i \left(\EV{ N_i}\right)^2$.
This observation allows us to set a simple bound for the values that $I(\rho)$ can take, when comparing different optical states defined over $m$ modes having the same overall photon number $\bar N_{tot} = \sum_i \EV{ N_i}$. 
It is straightforward to show that the minimum value for the invariant $I(\rho)$ is achieved when we have $m$ principal modes that are equally populated: 
\begin{equation}
\left.I(\rho)\right|_{min} = \frac{\bar N_{tot}^2}{m},
\end{equation}
while the maximum value is achieved in the case of all photons populating a single principal mode:
\begin{equation}
\left.I(\rho)\right|_{max} = \bar N_{tot}^2.
\end{equation}

We note that the eigenvalues of the coherency matrix are naturally conserved upon changes of basis and are therefore invariants in linear-optics transformations. However, not all the possible invariants derived as functions of $\EV{O_i}$ are independent. It can be shown, as described in detail in the Supplementary Material, that the maximal number of independent invariants which can be obtained is given by $m$. 
Thus, for a given number of modes $m$ and a given initial state preparation it is equivalent to check for the invariance of the $m$ eigenvalues of $\mathbf \Gamma^{(1)}$ or, for instance, of the quantity $I(\rho)$ plus other $m-1$ independent invariant quantities. 
Indeed, we show in the Supplementary Material that the information contained in the $m$ eigenvalues of $\mathbf \Gamma^{(1)}$ is equivalent to the information provided by the spectrum of the Hermitian operator $\rho_T$ defined in Eq.~\eqref{eq:dm_invariant}.

As mentioned in ~\cite{invariants}, one may not be limited to the construction of functions of $\EV{O_i}$, but may exploit also products of operators $O_i$, which potentially allows to increase the number of independent invariants which can be derived. 
Note that any product of operators $O_i$ can be cast into a linear combination of products of creation and annihilation operators, in equal number, possibly on different modes. Interestingly, one can show that such products are related to $n$-th order quantum correlation functions, on a set of modes specified by the indices $(s_1, \ldots,s_{n},s_{n+1},\ldots, s_{2n})$, which can be written in adimensional form as:
\begin{equation}
\Gamma^{(n)}(s_1,\ldots,s_{2n}) = \EV{ a^\dagger_{s_{1}} \ldots a^\dagger_{s_{n}}  a_{s_{n+1}} \ldots  a_{s_{2n}}}.
\end{equation}
Thus, our discussion on the invariants related to the quantum coherency matrix appears to be a particular case of a more general physical object, in which one considers a $2n$-order tensor having as entries the $n$-th order coherency matrix $\Gamma^{(n)}(s_1,\ldots,s_{2n})$, as discussed in the Supplementary Material.

However, the measurement of generic higher-order correlation functions (or related observables) presents increasing experimental challenges, due both to the growing complexity of the experimental setup and the exponential increase in the number of possible measurements—as the components of a $2n$-order tensor scale with $m^{2n}$. For this reason, in the present work we focus on the experimental investigation of invariants associated with first-order correlations.

\section{Experimental measurements and validation of Lie algebraic observables}

As described above, the evolution of a photonic state within a linear optical network implies the conservation of specific quantities, which can be derived from an underlying Lie algebraic structure. Moreover, such conservation laws are intrinsically connected to the concept of quantum optical coherency matrices. We now describe how the conservation of the described Lie invariant quantities for $n$-photon states evolving through a $m$-mode linear optical network has been experimentally verified, by carrying out observations featuring linear optical evolutions of up to $n=4$ photon states subjected to up to a $m=4$ mode unitary evolution.

\subsection{Experimental setup and results}
\begin{figure*}[ht]
    \centering
    \includegraphics[width = 0.999\textwidth]{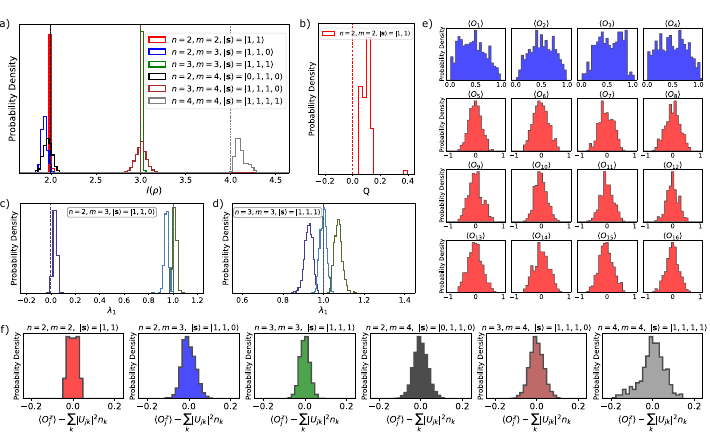}
    \caption{\textbf{Experimental distribution of the measured Lie algebraic quantities:}  
    (a) Invariance of the quantity $I(\rho)$, measured at the output of a $m\times m$ Haar-random linear optical evolution, with a given Fock state $\ket{\bm{s}}$ featuring $n$ photons distributed over $m$ modes.  
    (b-d) Spectral properties of the density operator $\rho_T$ as defined in Eq. \eqref{eq:dm_invariant}. Specifically, in (b) we analyze the principal component $Q$ of the infinite-cardinality spectrum of the operator $\rho_T$ in the $m=2$ modes configuration; in (c-d) we show the eigenvalues $\{ \lambda_1^j\}$ of the first block diagonal element of $\rho_T$ in the $m=3$ configuration, with $n=\{2,3\}$ photons. {As detailed in the Supplementary Materials, these eigenvalues are exactly those of the coherency matrix $\mathbf \Gamma^{(1)}$.} 
    (e) Histograms of the measured expectation values of the Lie observables $\langle O_i\rangle_{\rho}$ for the configuration with $m=4$ and $n=2$.  We denote with different colors histograms for $\langle O_j^z\rangle_{\rho}$ (blue) and for $\langle O_{jk}^{x,y}\rangle_{\rho}$ (red). This shows that Lie algebraic invariance arises not at the level of the individual expectation values $\Tr(\rho O_i)$, but only when considering their complete combination $I(\rho)$ as defined in Eq.~\eqref{eq:invariant}.
    (f) Histograms of the quantity $\langle O_j^z\rangle_{\rho}-\sum_k\abs{U_{jk}}^2n_k$, obtained from experimental data across different configurations. Here, $U$ represents the probed unitary transformations, while $\{n_k\}$ denote the input state occupation numbers. For each color-coded configuration, the vertical lines indicate the theoretical expectations for the corresponding Lie invariant (Panel a) or spectral quantities (Panels b-d).}
    \label{fig: Distribution}
\end{figure*}
We employ an advanced hybrid photonic architecture tailored for photon-based multi-photon experiments \cite{rodari2024semi,rodari2024experimental,rodari2024polarization}, whose main components for single-photon generation, preparation and manipulation of multi-photon resources are depicted in Fig. \ref{fig:exp_scheme}. A Quantum Dot (QD) single-photon source \cite{senellart2017high,heindel2023quantum,somaschi2016near,ollivier2020reproducibility,senellart2017high} is paired with a time-to-spatial demultiplexer (DMX) to obtain a stream of multi-photon states \cite{rodari2024semi,rodari2024experimental,rodari2024polarization,pont2022high,Pont2022quantifying}. Linear optical evolution is obtained within a eight-mode fully reconfigurable photonic integrated circuit (PIC) fabricated by femtosecond laser waveguide writing in glass \cite{corrielli2021femtosecond}. The precise reconfigurability of the interferometer is obtained via calibration \cite{pentangelo2024high} of the thermo-optic shifters fabricated on the device's surface \cite{Albiero2022toward}. 
In the experiment, the PIC is set to implement a known unitary transformation $U$ followed by the required measurement operators $O_{ij}$.  
We test Lie algebraic invariants with unitary transformations requiring up to $m=4$ modes within our $M=8$ modes PIC. Thus, we employ the first $m$ layers of the device by setting the corresponding $m(m-1)$ phases to implement a random unitary transformation in $SU(m)$, following the scheme in \cite{clements2016optimal}. Given that only $m$ layers are needed to achieve this, we utilize the remaining $(M-m)$ layers to implement the transformations required in order to measure the expectation values of the observables $O_i$ of Fig.~\ref{fig:liescheme}. This is accomplished through the action of balanced beam splitters and cross beam splitters, swapping two spatial modes, as shown in Fig. \ref{fig:liescheme}-(c-d) for both the $m=3$ and $m=4$ configurations. To reconstruct the expectation values of the observables, we determine the probability amplitudes for all possible photon-number outcomes, including those where multiple photons exit the linear interferometer bunching in the same mode, with pseudo-photon-number-resolving techniques. 
Specifically, at each output mode of the linear interferometer we insert a suitably constructed cascade of balanced beam splitters, implemented both on-chip and externally via in-fiber BS. In such a way, we can probe the full output photon-number distribution by correcting the measured $n$-fold coincidence events with a numerical factor taking into account the conditional probability of successful detection of a bunching event and the detector efficiencies, as detailed in the Supplementary Material.


We performed measurements of the output photon number distributions for several different experimental configurations, in terms of both the number of photons $n$ injected in the integrated chip and the number of modes $m$, which determine the dimension of the addressed linear-optical unitary transformation. 
Moreover the initial state injected in the PIC, expressed in the Fock-basis space, features at most one photon per input mode. 
For each configuration $\{n,m\}$, we randomly extracted (with respect to the Haar-random measure) about one hundred linear optical unitary transformation which were mapped onto phase settings to program the PIC via the so-called square decomposition algorithm \cite{clements2016optimal}.

\subsection{Invariance of the algebraic quantity $I(\rho)$}

We first tested the Lie algebraic invariant of Eq. \eqref{eq:invariant} across various Boson Sampling configurations, with varying number of photons $n$ and modes $m$, up to $n=m=4$ (see Fig.~\ref{fig: Distribution}. In each of them, we chose different no-collision Fock state inputs in the form $\ket{1^{\otimes n}0^{ \otimes (m-n) }}$; while implementing Haar-random unitary transformations $SU(m)$. 
The expectation values of the observables in Eq.~\eqref{eq:observables} were derived using the reconstructed populations of the output state, as detailed in the Supplementary Material. From them, the quantity $I(\rho)$ can be computed. The distribution of the experimentally probed values of the quantity $I(\rho)$, amongst a large number of Haar-random unitary evolutions and for different numbers of photons $n$ and modes $m$, are shown in Fig. \ref{fig: Distribution}-(a).
Furthermore, to evaluate the accuracy of the population reconstruction, we performed numerical simulations of the experiments: for a given output state $\rho$, we employ the metric $1-\text{TD}_\text{sim}$, where $\text{TD}_\text{sim}$ denotes the trace distance from the corresponding simulated output state $\rho_\text{sim}$, calculated as $\text{TD}_\text{sim} = \frac{1}{2}\mathrm{Tr}\abs{\rho-\rho_\text{sim}}$.
In Tab.~\ref{tab:stats} we present the averaged values of $1-\text{TD}_\text{sim}$, across all unitary transformations and observables. Additionally, Tab.~\ref{tab:stats} displays the average measured value of $I(\rho)$ and its associated standard deviation computed across all the dialled unitary evolution for all $\{n,m\}$ configurations. Note that, in each scenario, we obtain average values of $I_\text{exp}(\rho)$ which are compatible, within two standard deviations, with the theoretically expected value being $I(\rho) = n$, as proven in the Supplementary Material. 
We also note that a small deviation between the experimentally measured values of $I_\text{exp}(\rho)$ and the theoretically expected one can be explained by taking into account typical experimental imperfections of the apparatus, leading to small errors in the implementation of the observables for the Lie invariant measurement, or a miscalibration of the pseudo photon number resolving setup (see Supplementary Material for a qualitative numerical analysis of such effects).

Interestingly, the expectation value of the photon number operator in a given mode after a linear optical evolution $\tilde U$ satisfies the relation $\langle n_j \rangle = \sum_{k}|\tilde{U}_{jk}|^2 n_k$, where $n_k$ is the photon number in the j-th input of the interferometer \cite{PhysRevA.83.062307}, as discussed in the Supplementary Material. 
Since the expectation values of all observables are reconstructed from measured values of $\langle n_j \rangle$, all the measured quantities are a function of initial occupation numbers and unitary moduli of the optical evolution matrix, and thus they are not dependent on multi-photon distinguishability. However, as shown in Fig. \ref{fig: Distribution}-(e), each single expectation value $\expval{O_i}$ does distribute in the whole range $[-1,+1]$. Only by combining them via Eq. \eqref{eq:invariant}, a Lie algebraic invariant is obtained. 
Finally, Fig. \ref{fig: Distribution}-(f) reports histograms for the measured distributions of the quantity $\expval{O_j^z}_{\rho} - \sum_{k}|U_{jk}|^2 n_k$. There, in each configuration, the distributions are centered around zero, confirming the theoretical expectation $\expval{O_j^z}_{\rho} =  \expval{n_j}_{\rho} = \sum_{k}|U_{jk}|^2 n_k$. This observation indirectly confirms both that a given transformation is implemented on the device with a high-degree of accuracy, and that the cross configurations required to implement mode-swapping in the measurement stage are well implemented. 

\begin{table}[ht!]
\centering
\begin{tabular}{|c|c|c|c|c|c|c|}
\hline
$(n,m)$ & $\ket{\bm{s}}$  & ${I(\rho)}$ & Q / $\{\lambda_1^j\}$ & $\# U$ & $1-\text{TD}_\text{sim}$ \\
\hline
(2,2) & $\ket{1,1}$ & 2.008(40) & $0.010(38)$ & 1000 & 0.92(5) \\
\hline
(2,3)  & $\ket{1,1,0}$  & 1.936(73) & $\mqty(0.035(29) \\ 0.947(24) \\ 1.018(32))$ & 500 & 0.91(4) \\
\hline
(2,4)  & $\ket{0,1,1,0}$  & 1.970(57) & - & 400 & 0.90(4) \\ 
\hline 
(3,3)  & $\ket{1,1,1}$  & 3.012(6)  & $\mqty(0.927(20)\\ 1.000(16)\\ 1.073(19))$ & 1000 & 0.87(5) \\ 
\hline
(3,4)  & $\ket{1,1,1,0}$  & 3.007(65) & - & 1250 & 0.85(4) \\ 
\hline
(4,4)  & $\ket{1,1,1,1}$  & 4.111(58) & - & 120 & 0.81(3) \\ 
\hline
\end{tabular}

\caption{\textbf{Measured values for the Lie invariant $I(\rho)$}, obtained with different configurations corresponding to different inputs $\ket{\bm{s}}$ with a varying number of photons $n$ and modes $m$. For the $m=2$ and $m=3$ scenarios we report also the measured quantities related to the spectrum of $\rho_T$, which are $Q$ and $\{\lambda_{1}^j\}$ respectively.
In each row we provide the average and associated standard deviation of each quantity, computed over the total number $\# U$ of dialed linear unitary evolution. We also report the average value of $1-\text{TD}_\text{sim}$, across all unitary transformations and observables for the given configuration, as a measure of the quality of the population reconstruction.}
\label{tab:stats}
\end{table}

\subsection{Inference of the spectra of operator $\rho_T$}

As a final step we estimate, from the measured experimental data, the spectrum of the density operator as defined in Eq.~\eqref{eq:dm_invariant}. In the two-mode scenario, it is possible to show (see Supplementary Material) that the spectrum of $\rho_T$ is given by $\lambda_n^j = \frac{n}{2}(N_1+N_2) + j\sqrt{(N_1-N_2)^2+4\abs{R_{12}}^2}$. Here, $j \in \{-\frac{n}{2}, -\frac{n}{2} + 1, \dots, \frac{n}{2}-1, \frac{n}{2}\}$; $N_j = \langle n_j \rangle_{\rho}$ and $R_{jk} = \frac{1}{\sqrt{2}} \left( \langle O^x_{jk} \rangle_{\rho} + i \langle O^y_{jk} \rangle_{\rho} \right)$.
In order to test whether the spectrum of $\rho_T$ is invariant under linear optical evolution, we consider the input Fock state $\ket{\bm{s}} = \ket{1,1}$. In the regime with post-selected output the total number of photons $N_1 + N_2$ is conserved, so that the only non-trivial part of the spectrum is given by $Q = \sqrt{(N_1-N_2)^2+4\abs{R_{12}}^2}$. In the Supplementary Material we show that this quantity is related to the norm of the traceless part of the coherency matrix $\mathbf \Gamma^{(1)}$. Moreover we show that it is also related to the invariant $I(\rho)$ through the relation $Q^2 = 2I(\rho)-(N_1+N_2)^2$. This redundancy is expected since if we exclude the total number of photons we have only $m-1=1$ non trivial independent quantity in the $m=2$ scenario.
Fig.~\ref{fig: Distribution}-(b) shows the distribution of the values of $Q$ as obtained from the experimentally measured $N_j$ and $R_{12}$ for $n=m=2$, which is in agreement with the theoretically expected value $Q_\text{theo} = 0$. 
In the three-mode scenario, the analysis of the eigenvalue problem for the operator $\rho_T$ was restricted to its first block diagonal element, associated with $n=1$, as detailed in the Supplementary Material. There, we show that the spectra of such sub-element is given by the eigenvalues of a $3 \times 3$ Hermitian matrix $M$ which is found to be exactly the complex conjugate of the coherency matrix $\mathbf \Gamma^{(1)}$ (see Supplementary Material), and thus has the same eigenvalues. This matrix can be constructed via a measurement of $N_j$ and $R_{jk}$ and diagonalized numerically. In Fig.~\ref{fig: Distribution} (c-d), we show the distributions for the numerically computed eigenvalues $\{ \lambda_1^j\}$ of the matrix $M$ - reconstructed from the experimental data for each unitary $U$ - for two input state configurations $\ket{1,1,0}$ and $\ket{1,1,1}$, showing a very good agreement with their expected theoretical values denoted with vertical lines.


\section{Discussion and outlook}

In this work we have highlighted the relevance of Lie algebraic invariants in the characterization of the evolution of quantum states within linear optical systems, by means of Boson Sampling experiments carried out in a wide range of optical configurations. 
Leveraging a state-of-the-art hybrid photonic platform, we have experimentally validated the invariance of several different observable quantities that characterize the linear optical dynamic of photonic states, and that are derived from the Lie algebra of the linear-optics Hamiltonians.
Recalling the formalism of quantum correlation functions, we have shown how these invariants can be related to the conservation of fundamental coherence properties of the state, upon being subject to a linear optical transformation.
Our experimental results provide a concrete demonstration that the Lie algebraic framework — which could be regarded purely as an abstract theoretical tool — can be effectively harnessed in real-world photonic quantum information systems. Indeed, we have shown that the Lie algebraic framework - going beyond the typical formalism of Boson Sampling based on the scattering matrices - can be practically applied within a prototypical architecture for photonic quantum computation.
This not only offers a new lens through which to analyze quantum optical processes but could also lay the groundwork for the development of novel benchmarking protocol, tailored to test the linearity of an underlying bosonic dynamic or for the analysis of photon-based quantum information protocols.

\section*{Acknowledgements}

This work is supported by the ERC Advanced Grant QU-BOSS (QUantum advantage via non-linear BOSon Sampling, Grant Agreement No. 884676), the PNRR MUR project PE0000023-NQSTI (Spoke 4 and Spoke 7) and the European Union’s Horizon Europe research and innovation program under EPIQUE Project (Grant Agreement No. 101135288).
Fabrication of the fully reconfigurable photonic circuit was partially performed at PoliFAB (https://www.polifab.polimi.it/), the micro- and nano-fabrication facility of Politecnico di Milano. F.C. and R.O. would like to thank the PoliFAB staff for valuable technical support.

\bibliography{biblio_arxiv.bib}
\bibliographystyle{apsrev4-2}

\end{document}


\title{Supplementary Material: Observation of Lie algebraic invariants in Quantum Linear Optics}

\author{Giovanni Rodari}
\affiliation{Dipartimento di Fisica, Sapienza Universit\`{a} di Roma, Piazzale Aldo Moro 5, I-00185 Roma, Italy}

\author{Tommaso Francalanci}
\affiliation{Dipartimento di Fisica, Sapienza Universit\`{a} di Roma, Piazzale Aldo Moro 5, I-00185 Roma, Italy}

\author{Eugenio Caruccio}
\affiliation{Dipartimento di Fisica, Sapienza Universit\`{a} di Roma, Piazzale Aldo Moro 5, I-00185 Roma, Italy}

\author{Francesco Hoch}
\affiliation{Dipartimento di Fisica, Sapienza Universit\`{a} di Roma, Piazzale Aldo Moro 5, I-00185 Roma, Italy}

\author{Gonzalo Carvacho}
\affiliation{Dipartimento di Fisica, Sapienza Universit\`{a} di Roma, Piazzale Aldo Moro 5, I-00185 Roma, Italy}

\author{Taira Giordani}
\affiliation{Dipartimento di Fisica, Sapienza Universit\`{a} di Roma, Piazzale Aldo Moro 5, I-00185 Roma, Italy}

\author{Nicol\`o Spagnolo}
\affiliation{Dipartimento di Fisica, Sapienza Universit\`{a} di Roma, Piazzale Aldo Moro 5, I-00185 Roma, Italy}

\author{Riccardo Albiero}
\affiliation{Istituto di Fotonica e Nanotecnologie, Consiglio Nazionale delle Ricerche (IFN-CNR), Piazza Leonardo da Vinci, 32, I-20133 Milano, Italy}

\author{Niki Di Giano}
\affiliation{Istituto di Fotonica e Nanotecnologie, Consiglio Nazionale delle Ricerche (IFN-CNR), 
Piazza Leonardo da Vinci, 32, I-20133 Milano, Italy}
\affiliation{Dipartimento di Fisica - Politecnico di Milano, Piazza Leonardo da Vinci 32, 20133 Milano, Italy}

\author{Francesco Ceccarelli}
\affiliation{Istituto di Fotonica e Nanotecnologie, Consiglio Nazionale delle Ricerche (IFN-CNR), Piazza Leonardo da Vinci, 32, I-20133 Milano, Italy}

\author{Giacomo Corrielli}
\affiliation{Istituto di Fotonica e Nanotecnologie, Consiglio Nazionale delle Ricerche (IFN-CNR), Piazza Leonardo da Vinci, 32, I-20133 Milano, Italy}

\author{Andrea Crespi}
\affiliation{Istituto di Fotonica e Nanotecnologie, Consiglio Nazionale delle Ricerche (IFN-CNR), Piazza Leonardo da Vinci, 32, I-20133 Milano, Italy}
\affiliation{Dipartimento di Fisica - Politecnico di Milano, Piazza Leonardo da Vinci 32, 20133 Milano, Italy}

\author{Roberto Osellame}
\affiliation{Istituto di Fotonica e Nanotecnologie, Consiglio Nazionale delle Ricerche (IFN-CNR), Piazza Leonardo da Vinci, 32, I-20133 Milano, Italy}

\author{Ulysse Chabaud}
\affiliation{DIENS, \'Ecole Normale Sup\'erieure, PSL University, CNRS, INRIA, 45 rue d’Ulm, Paris, 75005, France}

\author{Fabio Sciarrino}
\email{fabio.sciarrino@uniroma1.it}
\affiliation{Dipartimento di Fisica, Sapienza Universit\`{a} di Roma, Piazzale Aldo Moro 5, I-00185 Roma, Italy}

\maketitle

\section{Single-photon source and multi-photon resource preparation}

Our experimental setup comprises a state-of-the-art hybrid photonic architecture tailored for multi-photon experiments \cite{rodari2024semi,rodari2024experimental,Hoch2025}. In particular, a hybrid approach is adopted, interconnecting a diverse array of technologies related to photon generation, manipulation and detection. 
As shown in Fig. 2 of the Main Text, the experimental apparatus consists of three subsequent stages.
Namely, we employ a single-photon source, based on Quantum Dot technology \cite{heindel2023quantum,senellart2017high}, excited at a repetition rate of $\sim 160$ MHz in the resonance fluorescence (RF) regime \cite{somaschi2016near,ollivier2020reproducibility,nowak2014deterministic,gazzano2013bright,senellart2017high}. The emitter itself is a single self-assembled InGaAs quantum dot embedded in an electrically controlled micro-cavity \cite{somaschi2016near}. The pulsed pump laser excites the source matching its excitonic transition energy, corresponding to an optical wavelength of approximately $\sim 928$ nm, and residual laser filtering is obtained via a set of HWP and QWP paired with a GTP in a standard cross-polarization scheme. 
The typical single-photon purity - measured via a Hanbury-Brown-Twiss setup - is around $\mathcal{P} = 1 - g^{(2)}(0) \sim 0.99$. Conversely, the pairwise indistinguishability amongst subsequently emitted single-photons - quantified via the Hong-Ou-Mandel visibility measured within a time-unbalanced Mach-Zehnder interferometer - is $V \sim 0.93$.

The QD source is interfaced with a time-to-space demultiplexing module (DMX) based on an acousto-optical modulator \cite{rodari2024semi,rodari2024experimental, pont2022quantifying,pont2022high} that converts the initial sequence of single photons, emitted at different times, in sets of photons distributed in several spatial modes, i.e., a multi-photon resource, which can be injected simultaneously in the input ports of the experiment. The pairwise indistinguishability between the photons comprising the four-photon state employed here is guaranteed by finely tuned free-space delay lines and polarization paddle controllers. The average visibility of the HOM of the three possible photon pairs among the three synchronized photons, estimated at the outputs of the integrated circuit, is in the range $\sim [0.89,0.94]$.

\section{Experimental Procedure to Measure the Expectation Value of Lie Observables from Output Probabilities}

In this section we describe the procedure used to reconstruct the expectation values of the observables needed for the evaluation of Lie invariant quantities.
We consider as a possible input state a Fock State $\ket{\bm{s}}$ of the form:
\begin{equation}
    \ket{\bm{s}} = \ket{n_1\dots n_m}, \quad \sum_{j=1}^m n_j = n.
\end{equation}
A generic output for for the state $\ket{\bm{s}}$ evolving through an $m$-mode interferometer is described by the state:
\begin{equation}\label{eq:psiout}
    \ket{\psi_{out}} = \sum_{k=1}^D \alpha_k \ket{\bm{t_k}},
\end{equation}
where $D = \binom{m+n-1}{n}$ is the dimension of the output space and $\{\ket{\bm{t_k}}\}$ is the set of Fock states $\ket{\bm{t_k}} = \ket{n_1^k\dots n_m^k}$ such that $\sum_{j=1}^m n_j^k = n$.

\medskip

In order to reconstruct Lie invariant quantities, it is required to evaluate the expectation values on the output state $\ket{\psi_{out}}$ of the observables associated to the following operators,
\begin{equation}\label{eq:observables}
\begin{cases}
& O_j^z = n_j = a^\dagger_j a_j \quad \text{for } j = 1, \dots, m, \\
& O_{jk}^x = \frac{1}{\sqrt{2}} 
\left( 
a^\dagger_j a_k + a^\dagger_k a_j 
\right) 
\quad \text{for } 1 \leq j < k \leq m, \\
& O_{jk}^y = \frac{i}{\sqrt{2}} 
\left( 
a^\dagger_j a_k - a^\dagger_k a_j 
\right) 
\quad \text{for } 1 \leq j < k \leq m.    
\end{cases}
\end{equation}
Experimentally, the expectation values of the first $m$ observables on a given output $\ket{\psi_{out}}$ can be simply obtained from the output probabilities $\{\abs{\alpha_k}^2\}$ as follows:
\begin{equation}\label{eq:n_out}
   \langle{O_j^{z}}\rangle_{\ket{\psi_{out}}} = \langle{n_j}\rangle_{\ket{\psi_{out}}} = \sum_{k=1}^{D} \abs{\alpha_k}^2  \langle{n_j}\rangle_{\ket{\bm{t_k}}}.
\end{equation}
The expectation values of the other observables are obtained by considering the following unitary operators acting on modes $j$ and $k$:
\begin{equation}
    \mathbf{T_x} \doteq \frac{1}{\sqrt{2}}\begin{bmatrix}
1 & 1 \\
1 & -1
\end{bmatrix} \qquad \mathbf{T_y} \doteq \frac{1}{\sqrt{2}}\begin{bmatrix}
i & 1 \\
i & -1
\end{bmatrix},
\end{equation}
which can be obtained from the action of beam splitters coupling modes $j$ and $k$ with $R=1/2$ and with $\phi_E = 0$ and $\phi_E = \frac{\pi}{2}$ respectively (where $R$ is the beam splitter reflectivity and $\phi_E$ is the beam splitter external phase).
The expectation values of the observables are then given by \cite{campos_beamsplitter}:
\begin{equation}\label{eq:O_xy}
\begin{aligned}
      & \langle{O_{jk}^{x}}\rangle_{\ket{\psi_{out}}} = \langle{\frac{1}{\sqrt{2}} 
\left( 
a^\dagger_j a_k + a^\dagger_k a_j 
\right)}\rangle_{\ket{\psi_{out}}} = \frac{1}{\sqrt{2}}\left(\langle{n_j}\rangle_{\hat{T}_x\ket{\psi_{out}}}-\langle{n_k}\rangle_{\hat{T}_x\ket{\psi_{out}}}\right),\\
 &   \langle{O_{jk}^{y}}\rangle_{\ket{\psi_{out}}} = \langle{\frac{i}{\sqrt{2}}
\left( 
a^\dagger_j a_k - a^\dagger_k a_j 
\right)}\rangle_{\ket{\psi_{out}}} = \frac{1}{\sqrt{2}}\left(\langle{n_j}\rangle_{\hat{T}_y\ket{\psi_{out}}}-\langle{n_k}\rangle_{\hat{T}_y\ket{\psi_{out}}}\right),
\end{aligned}
\end{equation}
where $\hat{T}_x$ and $\hat{T}_y$ denote the unitary operators acting on the output state corresponding to the mode transformations described by $\mathbf{T_x}$ and $\mathbf{T_y}$. The expectation values for the number operators in each mode are calculated again from the output probabilities, as in Eq. \eqref{eq:n_out}. 

\section{Data Analysis}

\begin{figure}[ht!]
    \centering
    \includegraphics[width = 0.75\columnwidth]{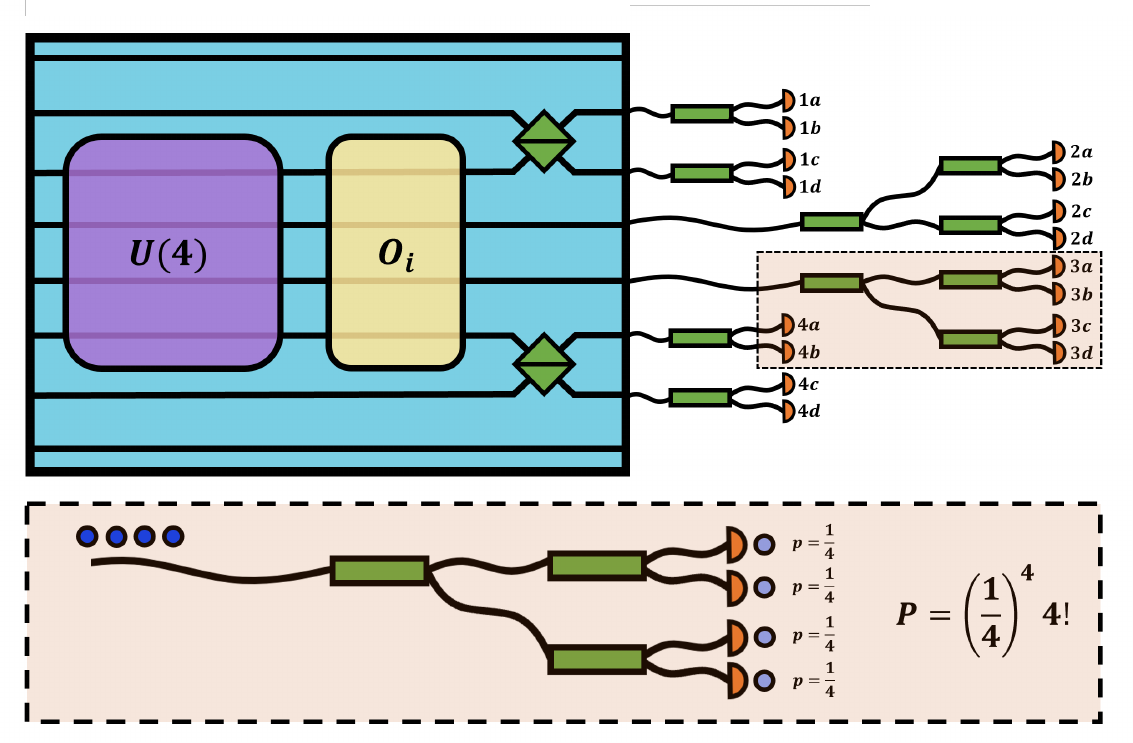}
    \caption{Schematic representation of the pseudo-number-resolving detection process, illustrated in the $m = 4$ case. The multi-photon states undergoes a linear optical dynamics described by the unitary evolution $U$. Then, the chip is setup to be able to measure the expectation value of the operator $O_i$. In order to do so, photons are detected using a pseudo-number resolving technique, from which photon-number distributions in each mode are reconstructed. Each output is split into auxiliary channels via a cascade of balanced beam splitters, implemented both in fiber and on-chip, as in the case of modes 1 and 4. 
    In each auxiliary mode, single photons are detected using SNSPD threshold detectors. The lower panel illustrates the scenario where four photons enter a cascade of perfectly balanced beam splitters, with each photon having a uniform probability $p = \frac{1}{4}$ of being transmitted to one of the four channels. This results in an overall probability $P = \left(\frac{1}{4}\right)^4 4!$ for this specific detection event to occur.
}
    \label{fig: number resolving}
\end{figure}

In this section, we describe the data analysis procedure that we followed to reconstruct the output probability amplitudes $\abs{\alpha_k}^2$ corresponding to each possible outcome $\ket{\bm{t}_k}$ for the $n$-photon, $m$-mode output state $\ket{\psi_{out}}$ introduced in Eq. \eqref{eq:psiout}. 
To achieve pseudo-number-resolving detection, we employ a probabilistic photon-number detection approach using auxiliary modes. Specifically, to resolve up to 4 photons in any output mode $j$ ($j = 1, 2, \dots, m$) with threshold detectors, each mode is divided among four auxiliary modes $j_\alpha$ ($\alpha = a, b, c, d$) via a cascade of beam splitters. In our experiment, this balanced splitting is implemented using both on-chip and external in-fiber single-mode 50:50 beam splitters (see Fig. \ref{fig: number resolving}). This setup allows us to probabilistically resolve up to four photons in each mode by combining the detection events across the four auxiliary modes. First, we reconstruct the set of $n$-tuples corresponding to all possible events where $n$ photons end up in $n$ different detectors and collect the relative photon counts. Then, we reconstruct an array of efficiency factors $\bm{\eta}$, with one efficiency factor assigned to each detector. Using this array, we correct the photon counts for each $n$-tuple by multiplying them by the efficiency factors of the corresponding detectors, thereby obtaining the effective photon counts.

Next, to reconstruct events where $n$ photons are distributed across multiple modes, we consider all possible $n$-tuple combinations of auxiliary mode detections leading to the event. For instance, in the case of $n=4$, $m=4$ and $\ket{\bm{t_k}} = \ket{3,1,0,0}$, where 3 photons end up in mode 1 and 1 photon in mode 2, we sum over all combinations of 3 photons detected in 3 different auxiliary modes of mode 1 (e.g., $1a, 1b, 1c$ or $1a, 1b, 1d$, etc.) paired with a single photon detected in an auxiliary mode of mode 2 (e.g., $2a$, $2b$, $2c$, or $2d$). The summed counts are then corrected by a factor $F$, which accounts for the probability of each sub-event detection.

A different correction factor is computed for each possible output state $\ket{\bm{t}_k}$, with corresponding occupation numbers $(n_1, \dots, n_m)$, with $n_j$ representing the number of photons in mode $j$. This is done by considering the probability $P_s$ of each sub-event leading to $\ket{\bm{t}_k}$. The balanced setup gives rise to an equal probability for each such combination, given by:
\begin{equation}
P_s = \bigg(\frac{1}{4}\bigg)^n \prod_{j=1}^m n_j!,
\end{equation}
where $P_s$ accounts for the probability of detecting a single photon in one of the auxiliary modes $j_\alpha$ after the beam splitter cascade, given by $P_{j \to j_\alpha} = 1/4$, and includes a factor $\prod_{j=1}^m n_j!$ to account for identical permutations of photons. The correction factor $F$ is then given by the inverse sum of the probabilities of all possible sub-events:
\begin{equation}
F = \bigg{(}\sum_{s} P_s\bigg{)}^{-1} = \bigg(\bigg(\frac{1}{4}\bigg)^n \prod_{j=1}^m \binom{4}{n_j} n_j!\bigg)^{-1}.
\end{equation}
Where we exploited the fact that the total number of equivalent sub-events is $\prod_{j=1}^m \binom{4}{n_j}$.

As a last step the corrected counts corresponding to each outcome $\ket{\bm{t}_k}$ are normalized to their sum, leading to the estimated probabilities $\abs{\alpha_k}^2$.

\section{Additional Insights on the Lie Algebra formalism applied to optical systems}

\subsection{Further analysis on the behaviour of the observables $\EV{O_i}$}

In this section, we give some further insights on the measured expectation values of the Lie observables defined in Eq.~\eqref{eq:observables}, given by $\langle O_i\rangle_{\rho} = \text{Tr}[O_i \rho]$. As discussed in the Supplementary Note 2, these expectation values are reconstructed by measuring the photon number expectation values $\langle n_j \rangle$ for each mode (Eqs.~\eqref{eq:n_out} and \eqref{eq:O_xy}). Notably, the independence of these expectation values from photon distinguishability is well established in the context of boson sampling and quantum walks \cite{PhysRevA.83.062307} and can be readily demonstrated using the Heisenberg picture. Specifically, considering a unitary matrix $U$ that describes the transformation of the creation operators under linear optical evolution,  
\begin{equation}
    a'^\dagger_j = \sum_{k=1}^m U_{jk} a^\dagger_k, 
\end{equation}  
the expectation value of the photon number in mode $j$ evolves as  
\begin{equation}\label{eq: n_prime}
    \langle n'_j \rangle = \langle a'^\dagger_j a'_j \rangle = \sum_{k=1}^m \sum_{l=1}^m U_{jk} U^*_{jl} \langle a^\dagger_k a_l \rangle = \sum_{j=1}^m U_{jk} U^*_{jk} \langle a^\dagger_k a_k \rangle =\sum_{k=1}^m \left| U_{jk} \right|^2 \langle n_k \rangle.
\end{equation}  
Here, we have used the fact that $\langle a^\dagger_k a_l \rangle = \delta_{kl} \langle a^\dagger_k a_k \rangle$. 
As already stated in the main text, this expression explicitly shows that the expectation value of the photon number in a given mode is independent of photon indistingushability, as it depends only on the square moduli of the unitary matrix elements and the initial expectation values. Note that for an input Fock state $\ket{\bm{s}} = \ket{n_1,..., n_m}$, these expectation values simply correspond to the initial occupation numbers $n_k$ in each mode.\\
Following the result of Eq.~\eqref{eq: n_prime} we provide some more details about the observables $O_i$ and their role in the evaluation of the invariant quantity:
\begin{equation}
    I(\rho) = \sum_{i=1}^{m^2} \Tr(O_i \rho)^2=\sum_{i=1}^{m^2}\langle{O_i}\rangle_{\rho}^2.
\end{equation}
The first $m$ observables are given by the mean number of photons per mode: $\langle O_j^z\rangle_{\rho} = \langle{n_j}\rangle_{\rho}$. It can be shown that after a linear optical evolution from an initial state $\ket{\bm{s}}$ to a final state $\ket{\psi_{out}}$, the mean number of photons in any mode $\langle{n_j}\rangle_{\ket{\psi_{out}}}$ has the following bounds:
\begin{equation}\label{eq:n_mean}
    \langle{n_j}\rangle_{\ket{\psi_{out}}} \geq \min_k\langle{n_k}\rangle_{\ket{\bm{s}}} \qquad \langle{n_j}\rangle_{\ket{\psi_{out}}} \leq \max_k\langle{n_k}\rangle_{\ket{\bm{s}}}.
\end{equation}
Indeed, using  Eq.~\eqref{eq: n_prime}  one has:
\begin{equation}
    \langle{n_j}\rangle_{\ket{\psi_{out}}} = \sum_{k=1}^m \left| U_{jk} \right|^2 \langle n_k \rangle_{\ket{\bm{s}}}
    \begin{cases}
    & \geq \min_k \langle n_k \rangle_{\ket{\bm{s}}} \sum_{k=1}^m \left| U_{jk} \right|^2 = \min_k \langle n_k \rangle_{\ket{\bm{s}}} \\
    & \leq \max_k \langle n_k \rangle_{\ket{\bm{s}}} \sum_{k=1}^m \left| U_{jk} \right|^2 = \max_k \langle n_k \rangle_{\ket{\bm{s}}},
    \end{cases}
\end{equation}
where we used the fact that for a unitary matrix \(U\), \(\sum_{k=1}^m \left| U_{jk} \right|^2 = 1\).

\begin{figure}[ht!]
    \centering
    \includegraphics[width = 0.95\columnwidth]{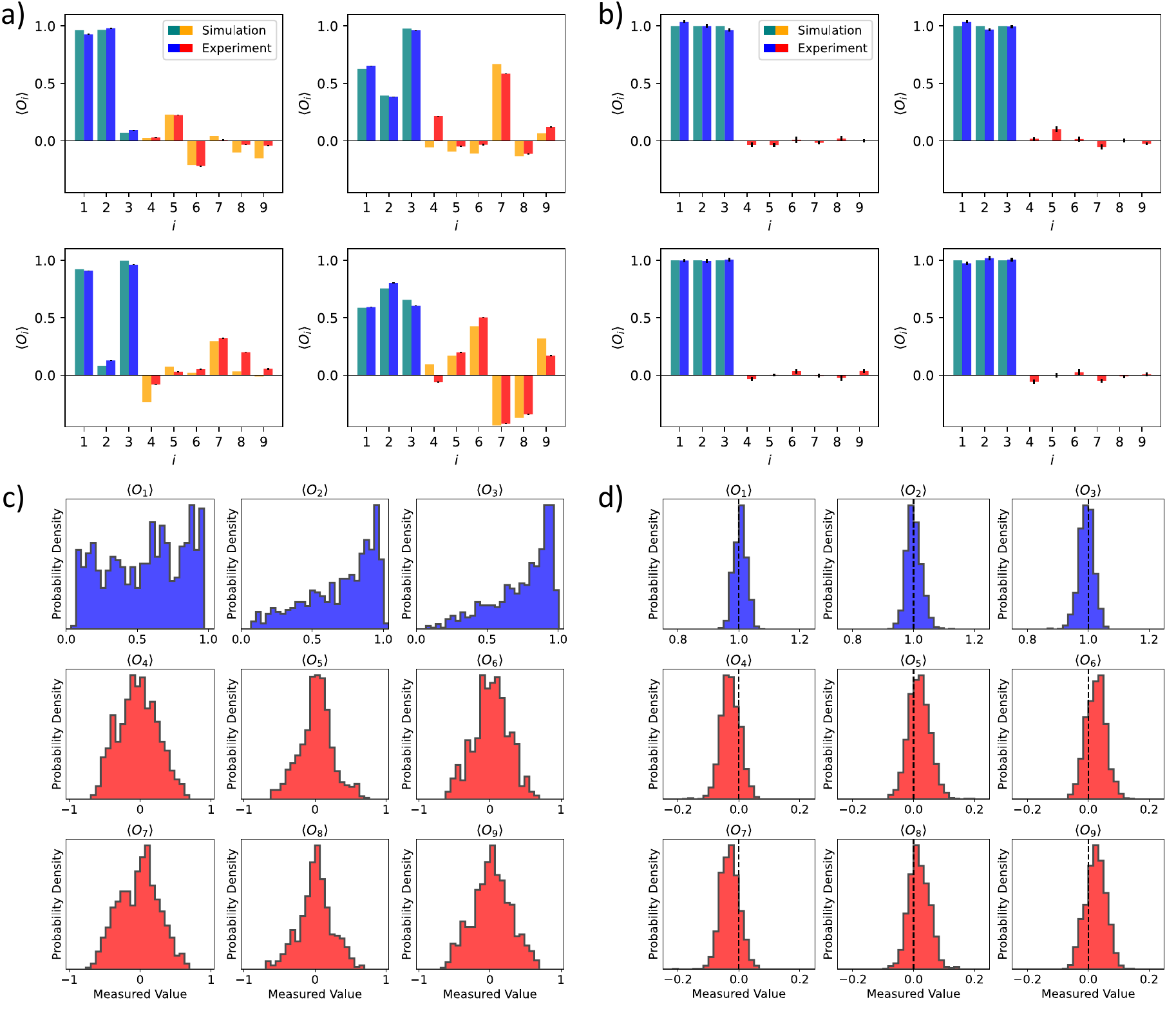}
    \caption{a-b) Example of expectation values of observables, $ \langle O_i \rangle $, in the $ m = 3 $ scenario, as measured after four different unitary transformations $U$. Error bars are obtained from uncertainties in single photon counts given by Poisson statistics. c-d) Histograms of the expectation values of the measured observables $O_i$ on the output states in the same configurations over all the different unitary transformations. a-c) correspond to input state $\ket{\bm{s}} = \ket{1,1,0} $ and b-d) to $ \ket{\bm{s}} = \ket{1,1,1} $. In all cases we highlight the measured values using different colors: the first $ m $ observables correspond to $ O_j^z = n_j $, while the remaining $ m(m-1) $ correspond to $ O_{jk}^x $ and $ O_{jk}^y $. As expected, the latter are nonzero for the input $ \ket{\bm{s}} = \ket{1,1,0} $ but take values close to zero in the case of $ \ket{\bm{s}} = \ket{1,1,1} $. }
    \label{fig: observables}
\end{figure}

In particular, for any input state of the form $\ket{\bm{s}} =\ket{1,1,...,1}$, with one photon in each mode, it follows from Eq.~\eqref{eq:n_mean} that $\langle{n_j}\rangle_{\ket{\psi_{out}}} =1$ for each $k = 1,\dots,m$. As a consequence we have:
\begin{equation}
    \sum_{i=1}^m\langle{O_j^z}\rangle_{\ket{\psi_{out}}}^2 = \sum_{j=1}^m\langle{n_j}\rangle_{\ket{\psi_{out}}}^2 = n.
\end{equation}
Note that for this kind of input state we have:
\begin{equation}
    I(\ket{\bm{s}}\bra{\bm{s}}) = \sum_{j=1}^m\langle n_j \rangle_{\ket{\bm{s}}}^2 = n,
\end{equation}
so that for the invariant to be conserved the expectation values of the other observables ($O_{jk}^x$ and $O_{jk}^y$) on the output state have to be zero. Remarkably, in this case the conservation of the invariant follows from the conservation of the expectation values of the observables themselves. We note, however, that in general this is not the case: if we consider input state different from $\ket{\bm{s}} = \ket{1,1,\dots,1}$ it is possible to have $\sum_{j=1}^m \langle{n_j}\rangle_{\ket{\psi_{out}}}^2 \neq n$, and thus the expectation values of the other observables ($O_{jk}^x$ and $O_{jk}^y$) on the output state can be different from zero. In Fig. \ref{fig: observables} we show some examples of measured expectation values of the observables as obtained after the evolution of different input states in the $m=3$ scenario, as well as histograms of the measured values over all the different unitary evolutions.

\subsection{Number of invariants constructed with $\EV{O_i}$}

It is interesting to evaluate the number of independent scalar invariants that can be constructed using the expectation values of the observables $O_i$. Independent quantities cannot be, in general, retrieved from one another.

As discussed in the Main Text, the full set of the $\EV{O_i}_\rho$ carries the same information than the entries of the coherency matrix $\mathbf{\Gamma}^{(1)}$. Thus, the above question can be reformulated as evaluating the number of independent quantities characterizing the coherency matrix, that are invariant upon a change of basis. The $m$ eigenvalues $\lambda_j = \EV{N_j}$ of $\mathbf{\Gamma}^{(1)}$ are such a set of independent invariant quantities. 

Actually, they form a maximal set of independent invariant quantities. 
Indeed, let us assume \textit{ab absurdo} that there exists a further quantity $\lambda^*$ that characterizes the coherency matrix, which is independent from the set of $\lambda_j$ and is invariant by change of basis. It means that one can build two matrices $\mathbf{\Gamma}^{(1)}_a \neq \mathbf{\Gamma}^{(1)}_b$, with the same set of eigenvalues but one with $\lambda^*=\lambda_a$, and the other with $\lambda^*=\lambda_b$. As $\lambda^*$ is invariant by a change of basis, it should not be possible to transform $\mathbf{\Gamma}^{(1)}_a$ into $\mathbf{\Gamma}^{(1)}_b$ by a change of basis. But this is absurd as, actually, it is \textit{always} possible to transform one into the other, by a unitary change of basis, two Hermitian matrices that share the same set of eigenvalues.

In particular, the invariant $I(\rho)$ is a function of the eigenvalues of $\mathbf{\Gamma}^{(1)}$, hence it is not another independent quantity.  The information contained in all possible invariants that are functions of the $\EV{O_i}$ coincides with the information contained in the $m$ real eigenvalues of the coherency matrix $\mathbf{\Gamma}^{(1)}$, or with any other $m$ analogous quantities.

\section{Invariants related to higher-order quantum correlation functions}

\subsection{Tensorial description for higher-order correlations}

Functions of products of operators $O_i$ may be exploited to find other invariant quantities, potentially independent from the ones constructed with only the single-operator expectation values $\EV{O_i}$.
For instance, in Ref.~\cite{invariants} we find the example of the covariance matrix, with entries:
\begin{equation}
M_{ij} =\EV{O_i}\EV{O_j} - \frac{\EV{O_i O_j}+\EV{O_j O_i}}{2},
\label{eq:Mcov}
\end{equation}
whose eigenvalues are shown to be conserved upon linear-optical transformations.

As the individual $O_i$ are combinations of terms $a^\dagger_j a_k$, any product of $m$ operators of the kind $O_i$ will be a linear combination of products of creation and annihilation operators, possibly on different modes. Importantly, in each of the latter products, the number of annihilation operators is the same as the number of creation operators. 

In detail, right after the multiplication $O_i O_j \ldots$, the creation and annihilation operators might not be in normal order (all creation on the left and all annihilation on the right). By applying the commutators
$a_j a_k^\dagger - a_k^\dagger a_j = \delta_{jk}$
one then rewrites the products of $m$ creation and $m$ annihilation operators into linear combinations of other (normally ordered) products of creation and annihilation operators. Importantly, each term of this combination is a product of a certain number of creation operators and the same number of annihilation operators, such number being lower than or equal to $m$.  

The general $n$-th order quantum correlation function on a set of modes, specified by the indices $(s_1, \ldots,s_{n},$ $s_{n+1},\ldots, s_{2n})$, can be written in adimensional form as:
\begin{equation}
\Gamma^{(n)}(s_1,\ldots,s_{2n}) = \EV{ a^\dagger_{s_{1}} \ldots a^\dagger_{s_{n}}  a_{s_{n+1}} \ldots   a_{s_{2n}}}.
\end{equation}
Therefore, the expectation value of the product of the $O_i$ operators can be traced back to combinations of quantum correlation functions of different order, which also have interesting properties with respect to change of basis corresponding to linear optical transformations.

In fact, let:
\begin{align} \label{eq:transf}
b^\dagger_i &= \sum_{j=1}^m U_{ij} a^\dagger_j,  & b_i &= \sum_{j=1}^m U^*_{ij} a_j  =\sum_{j=1}^m (U^{-1})_{ji} a^\dagger_j
\end{align}
be the basis change induced by a linear-optics transformation on the set of modes. The correlation function $\Gamma^{(n)}(s_1,\ldots,s_{2n})$ transforms under this basis change as:
\begin{align}
\Gamma^{(n)}_b (s_1,\ldots,s_{2n}) &= \EV{ b^\dagger_{s_{1}} \ldots  b^\dagger_{s_{n}} b_{s_{n+1}} \ldots b_{s_{2n}}}\\ 
&=
\sum_{t_1=1}^m \ldots \sum_{t_{2n}=1}^m U_{s_1 t_1} \ldots U_{s_n t_n} \, U^*_{s_{n+1} t_{n+1}} \ldots U^*_{s_{2n} t_{2n}} \, \EV{ a^\dagger_{t_{1}} \ldots a^\dagger_{t_{n}}  a_{t_{n+1}} \ldots  a_{t_{2n}}} \\
&=\sum_{t_1=1}^m \ldots \sum_{t_{2n}=1}^m U_{s_1 t_1} \ldots U_{s_n t_n} \, (U^{-1})_{t_{n+1} s_{n+1} } \ldots (U^{-1})_{t_{2n} s_{2n}} \,\Gamma^{(n)}_a (t_1,\ldots,t_{2n}). \label{eq:tensorTransf}
\end{align}
One notes that the correlation functions of order $n$ are transforming themselves akin to the components of a $2n$-order $m$-dimensional tensor. 
Indeed, a tensor with contra-variant indices $t_1,\ldots, t_n$ and co-variant indices $t_{n+1}, \ldots, t_{2n}$ transforms as follows (using the customary index notation for tensors)  \cite{kay1988schaum}:
\begin{equation}
\bar{T}^{s_1,\ldots, s_n}_{s_{n+1}, \ldots, s_{2n}} = U^{s_1}_{t_1} \ldots  U^{s_n}_{t_n} \,(U^{-1})^{t_{n+1}}_{s_{n+1}} \ldots(U^{-1})^{t_{2n}}_{s_{2n}}\; T^{t_1,\ldots, t_n}_{t_{n+1}, \ldots, t_{2n}}\,, \label{eq:tensorTransfTheory}
\end{equation}
where $U$ is the Jacobian matrix of the coordinate transformation. Differently from the typical tensors defined on the real space, the elements $\Gamma^{(n)}(s_1,\ldots,s_{2n})$ are complex and  in the case of Eq.~\eqref {eq:tensorTransf} $U$ is a complex matrix.

In addition, the quantum correlation functions possess relevant symmetry properties already noted by Glauber \cite{glauber1963quantum}. For instance,
\begin{equation}
\Gamma^{(n)}(s_1,\ldots,s_n,s_{n+1},\ldots,s_{2n}) = (\Gamma^{(n)}(s_{2n},\ldots,s_{n+1},s_{n},\ldots,s_{1}))^*,\label{eq:tensorHerm}
\end{equation} 
and furthermore $\Gamma^{(n)}(s_1,\ldots,s_n,s_{n+1},\ldots,s_{2n})$ is unchanged for any permutation of the arguments $(s_1, \ldots, s_n)$ and of the arguments $(s_{n+1}, \ldots, s_{2n})$. Therefore, many of the elements of the tensor are not independent one from the other. 

A certain structure in the tensor components may also come from specific symmetries of the input light state. Interestingly, a dramatic simplification is observed in the case of a state representable as the excitation of a single principal mode, as in the latter case all the elements of the $2n$-th order tensor are proportional one to the other, namely only one element is independent \cite{titulaer1965correlation}. In the general case, however, one cannot rely on such additional relations.

The tensor-like nature of $\mathbf{\Gamma}^{(n)}$ warrants for the presence of scalar invariants with respect to the change of basis. In fact, one simple invariant of the tensor could be the (squared) Frobenius norm, in analogy with $I(\rho)$:
\begin{equation}
\left\lVert \mathbf{\Gamma}^{(n)} \right\rVert^2 = \sum_{s_1} \ldots \sum_{s_{2n}} \left|\Gamma^{(n)}(s_1,\ldots,s_{2n})\right|^2 .\label{eq:tensorNorm}
\end{equation}
Further invariants may be the eigenvalues of the tensor (in this regard, it should be noted that different definitions of eigenvalues exist for a tensor \cite{qi2007eigenvalues,cui2016eigenvalue}). We discuss in the next subsection the calculation of one kind of tensor eigenvalues, proving directly their invariance with respect to the change of basis in Eq.~\eqref{eq:transf}. 
In any case, retrieving $\left\lVert \mathbf{\Gamma}^{(n)} \right\rVert^2$ or finding the eigenvalues requires in principle to characterize all the (independent) entries of the tensor, which correspond to a large number of correlation functions for the possible different combinations $(s_1,\ldots,s_{2n})$.  This may pose relevant difficulties from an experimental point of view.

\subsection{Unfolded matrices and eigenvalues}

Even-order tensors can be associated to a square matrix $M$ with dimension $m^n \times m^n$, by adopting the unfolding presented e.g. in Ref.~\cite{cui2016eigenvalue}. In detail, the matrix unfolding of our tensor $\mathbf{\Gamma}^{(n)}$ is described by:
\begin{equation}
M^{(n)}_P(h,k) = \Gamma^{(n)} (s'_1,\ldots,s'_{n},s'_{n+1},\ldots,s'_{2n}),
\end{equation}
with
\begin{gather}
h = m^{n-1} (s'_1-1)+m^{n-2} (s'_2-1)+\ldots+m(s'_{n-1}-1)+s'_{n}\\
k = m^{n-1} (s'_{n+1}-1)+m^{n-2} (s'_{n+2}-1)+\ldots+m(s'_{2n-1}-1)+s'_{2n}.
\end{gather}
The mode indices $\lbrace s'_j\rbrace$ are related to the mode indices $\lbrace s_j \rbrace$ by means of a permutation matrix $P$, acting independently on the subset $(s_1,\ldots,s_n)$ and on the subset $(s_{n+1},\ldots,s_{2n})$ according to:
\begin{align}
(s'_1,\ldots,s'_n) &= (s_1,\ldots,s_n) P, & (s'_{n+1},\ldots,s'_{2n}) &= (s_{n+1},\ldots,s_{2n}) P.
\end{align}
For instance, if we keep the natural ordering (i.e.~we take $P$ as the identity), in the case of the second-order correlation tensor ($n=2$) for two modes ($m=2$)  we build the associated matrix:
\begin{equation}
\mathbf{M}^{(2)} = \matr{
\Gamma^{(2)}(1,1,1,1) & \Gamma^{(2)}(1,1,1,2) & \Gamma^{(2)}(1,1,2,1) & \Gamma^{(2)}(1,1,2,2)\\
\Gamma^{(2)}(1,2,1,1) & \Gamma^{(2)}(1,2,1,2) & \Gamma^{(2)}(1,2,2,1) & \Gamma^{(2)}(1,2,2,2)\\
\Gamma^{(2)}(2,1,1,1) & \Gamma^{(2)}(2,1,1,2) & \Gamma^{(2)}(2,1,2,1) & \Gamma^{(2)}(2,1,2,2)\\
\Gamma^{(2)}(2,2,1,1) & \Gamma^{(2)}(2,2,1,2) & \Gamma^{(2)}(2,2,2,1) & \Gamma^{(2)}(2,2,2,2)} .\label{eq:M2}
\end{equation}
Due to the relation \eqref{eq:tensorHerm} and the other symmetry properties of quantum correlation functions, the unfolded matrix $\mathbf{M}$ is Hermitian and thus possesses $m^n$ real eigenvalues.

We show in the following that a change of basis of the tensor $\mathbf{\Gamma}^{(n)}$ is associated to a change of basis of the matrix $\mathbf{M}^{(n)}$. In discussing the relation between the two transformations we will bring as examples the formulas for the case of $n=2$ and $m=2$ (namely the case of matrix $\mathbf{M}^{(2)}$ in Eq.~\eqref{eq:M2}). In fact, the procedure can be generalized to higher order and larger dimension. 

Let $\mathbf{U}$ be the $m \times m$ matrix associated to the basis change of the tensor $\mathbf{\Gamma}^{(n)}$ as in Eqs.~\eqref{eq:transf}. In the $n=m=2$ case, in particular: 
\begin{equation}
\mathbf{U}=\matr{U_{11} & U_{12}\\ U_{21} & U_{22}} .
\end{equation}
Let us now build a block-diagonal matrix $\mathbf{U}_B$ of dimension $m^{n} \times m^{n}$, by repeating $\mathbf{U}$ on the diagonal $m^{n-1}$ times:
\begin{equation}
\mathbf{U}_B = \matr{ \mathbf{U} & \mathbf{0}\\ \mathbf{0} & \mathbf{U}} = \matr{
U_{11} & U_{12} & 0 & 0\\
U_{21} & U_{22} & 0 & 0\\
0 & 0 & U_{11} & U_{12}\\
0 & 0 & U_{21} & U_{22}} .
\end{equation}
We study what happens when we apply to $\mathbf{M}^{(n)}$ a basis change as:
\begin{equation}
\bar{\mathbf{M}}^{(2)} = \mathbf{U}_B \mathbf{M}^{(2)} \mathbf{U}^\dagger_B .\label{eq:VB_M_VB}
\end{equation}
The action of $\mathbf{U}_B$ is perhaps better understood if we divide also the matrix $\mathbf{M}^{(2)}$ into blocks of size $m \times m$:
\begin{equation}
\mathbf{M}^{(2)} = \matr{ \mathbf{M}_{11} & \mathbf{M}_{12}\\ \mathbf{M}_{21} & \mathbf{M}_{22} } = \left[\begin{array}{cc|cc}
\Gamma^{(2)}(1,1,1,1) & \Gamma^{(2)}(1,1,1,2) & \Gamma^{(2)}(1,1,2,1) & \Gamma^{(2)}(1,1,2,2)\\
\Gamma^{(2)}(1,2,1,1) & \Gamma^{(2)}(1,2,1,2) & \Gamma^{(2)}(1,2,2,1) & \Gamma^{(2)}(1,2,2,2)\\ \hline
\Gamma^{(2)}(2,1,1,1) & \Gamma^{(2)}(2,1,1,2) & \Gamma^{(2)}(2,1,2,1) & \Gamma^{(2)}(2,1,2,2)\\
\Gamma^{(2)}(2,2,1,1) & \Gamma^{(2)}(2,2,1,2) & \Gamma^{(2)}(2,2,2,1) & \Gamma^{(2)}(2,2,2,2)\end{array}\right].
\end{equation}
We note that, inside a given block, the entries $\Gamma^{(2)}(s_1,\ldots,s_{n},s_{n+1},\ldots,s_{2n})$ change the indices $s_{n}$ (which basically indicates the row inside the block) and $s_{2n}$ (which basically indicates the column). The other indices $(s_1,\ldots,s_{n-1})$ and $(s_{n+1},\ldots,s_{2n-1})$ take instead constant values. 
The basis transformation of Eq.~\eqref{eq:VB_M_VB} is actually a basis transformation of each block through the matrix $\mathbf{U}$:
\begin{equation}
\bar{\mathbf{M}}^{(2)} = \mathbf{U}_B \mathbf{M}^{(2)} \mathbf{U}^\dagger_B = \matr{ \mathbf{U} \mathbf{M}_{11} \mathbf{U}^\dagger & \mathbf{U} \mathbf{M}_{12} \mathbf{U}^\dagger \\ \mathbf{U} \mathbf{M}_{21} \mathbf{U}^\dagger  & \mathbf{U} \mathbf{M}_{22} \mathbf{U}^\dagger  } ,
\end{equation}
and in detail the matrix elements are:
\begin{equation}
\bar{M}^{(2)}(h,k)= \sum_{t_n=1}^m \sum_{t_{2n}=1}^m U_{s_n\,t_n} (U^\dagger)_{t_{2n}\, s_{2n}} \Gamma^{(2)}(s_1,\ldots,s_{n-1},t_{n},s_{n+1},\ldots,t_{2n}).
\end{equation}

Thus, being $U^{-1} \equiv U^\dagger$, we are indeed beginning to build the transformation as in Eq.~\eqref{eq:tensorTransf}, but affecting only the indices $s_n$ and $s_{2n}$. To proceed, we can apply a suitable permutation matrix $\mathbf{P}_S$ that shifts cyclically the indices as:
\begin{align}
(s_1, \ldots, s_{n-1},s_{n}) &\rightarrow (s_2 \ldots s_{n}, s_1) & (s_{n+1}, \ldots, s_{2n-1},s_{2n}), &\rightarrow (s_{n+2} \ldots s_{2n}, s_{n+1}),
\end{align}
and repeat the basis change by means of $\mathbf{U}_B$:
\begin{equation}
\mathbf{U}_B \mathbf{P}_S \mathbf{U}_B \; \mathbf{M}^{(2)} \; \mathbf{U}^\dagger_B \mathbf{P}^\dagger_S \mathbf{U}^\dagger_B.
\end{equation}
This applies the basis change to the indices $s_{n-1}$ and $s_{2n-1}$ leaving unaltered the other ones. In case $n=2$, we may apply one last time the permutation and recover the correct index ordering. In general, we need overall to apply $n$ times the basis change $\mathbf{U}_B \ldots \mathbf{U}_B^\dagger$, alternated with the permutation $\mathbf{P}_S \ldots \mathbf{P}_S^\dagger$. The full transformation from $\mathbf{M}^{(2)}_a$ in the basis $a$ to $\mathbf{M}^{(2)}_b$ in the basis $b$ reads:
\begin{equation}
\mathbf{M}^{(2)}_b = \left(\underbrace{\mathbf{P}_S \mathbf{U}_B \ldots \mathbf{P}_S \mathbf{U}_B}_{n\;\text{times}} \right) \mathbf{M}^{(2)}_a \left(\underbrace{\mathbf{U}^\dagger_B \mathbf{P}_S \ldots \mathbf{U}^\dagger_B \mathbf{P}^\dagger_S }_{n\;\text{times}}  \right).
\end{equation}
This operation is indeed equivalent to a single basis change of the full matrix according to $\mathbf V = \underbrace{  \mathbf{P}_S \mathbf{U}_B \ldots  \mathbf{P}_S \mathbf{U}_B}_{n\;\text{times}}$:
\begin{equation}
\mathbf{M}^{(2)}_b  = \mathbf{V}\, \mathbf{M}_a^{(2)}\, \mathbf{V}^\dagger.
\end{equation}

Hence, a basis change $\mathbf{U}$ of the tensor corresponds to a basis change $\mathbf{V}$ of the unfolded matrix. This means that the eigenvalues of the matrix, which are invariant upon any change of basis of the matrix itself, are invariant also for those matrices $\mathbf{V}$ that represent a change of basis of the tensor. They can thus be considered as invariants of the tensor itself \cite{cui2016eigenvalue}.

\subsection{Projection of the density operator on subspaces related to high-order correlations}

From the previous discussion it is apparent that the correlation functions $\Gamma^{(n)}(s_1,\ldots,s_{2n})$ are expectation values of operators, that are normally ordered products of creation and annihilation operators in equal number, and that we may indicate as:
\begin{equation}
 G^{(n)}_{s_1,\ldots,s_{2n}} = a^\dagger_{s_{1}} \ldots a^\dagger_{s_{n}}  a_{s_{n+1}} \ldots a_{s_{2n}} .\label{eq:Goperator}
\end{equation}
These operators in Eq.~\eqref{eq:Goperator} and their linear combinations  form a subspace (for a given $n$). Note that the change of basis applies to the operators $ G^{(n)}$ precisely as in Eq.~\eqref{eq:tensorTransf}:
\begin{equation}
G^{(n)}_b (s_1,\ldots,s_{2n}) = \sum_{t_1=1}^m \ldots \sum_{t_{2n}=1}^m U_{s_1 t_1} \ldots U_{s_n t_n} \, (U^{-1})_{t_{n+1} s_{n+1} } \ldots (U^{-1})_{t_{2n} s_{2n}} \, G^{(n)}_a (t_1,\ldots,t_{2n}). \label{eq:tensorTransf2}
\end{equation}
Thus, the change of basis induced by a linear-optics transformation maintain a combination of $ G^{(n)}$ inside the subspace, analogously to what is observed for the operators $O_i$.

In fact, one could also define an orthogonal basis for the subspace of order $n$ in terms of Hermitian operators $ O^{(n)}_i$ which have several properties in common with the $O_i$ defined in Ref.~\cite{invariants} (which actually correspond to the case $n=1$). The $ O^{(n)}_i$ would be indeed linear combinations of the $ G^{(n)}_{s_1,\ldots,s_{2n}}$, and thus their expectation values $\EV{O^{(n)}_i}=\Tr(O^{(n)}_i \rho)$ would be linear combinations of the quantum correlation functions $\Gamma^{(n)}(s_1,\ldots,s_{2n})$. 

Quantum mechanical operators on the optical modes can be expressed, under rather general hypotheses, as ordered power series of the bosonic operators of the modes \cite{cahill1969ordered}. The density operator $\rho$ describing the state of the electromagnetic field could be also expressed according to this decomposition. Hence, the density operator $\rho$ may have a meaningful projection on the subspace of the operators $ G^{(n)}_{s_1,\ldots,s_{2n}}$:
\begin{equation}
 \rho_T^{(n)} = \sum_i \Tr(O^{(n)}_i \rho) \,  O^{(n)}_i.
\end{equation}
Again in analogy with the $\rho_T$ discussed in Ref.~\cite{invariants} and in other parts of this work, linear-optical transformations do not mix the projections $\rho_T^{(n)}$ of different order. In addition, as the coefficients $\Tr(O^{(n)}_i \rho)$ can in principle be written as linear combinations of quantum correlation functions $\Gamma^{(n)}(s_1,\ldots,s_{2n})$, the information content of the projection $\rho_T^{(n)}$ coincide with the information content of the tensor $\mathbf \Gamma^{(n)}$. This may connect possible invariant quantities related to $ \rho_T^{(n)}$ to the invariants that we have discussed above, related to the tensor $\mathbf \Gamma^{(n)}$.
In Supplementary Note 8, we actually show that this is the case for $\rho_T \equiv\rho_T^{(1)}$, the operator discussed in Ref.~\cite{invariants} and in the main text.

\section{Impact of experimental imperfections on the measurement of Lie observables}

\begin{figure}[ht!]
    \centering
    \includegraphics[width = 0.99\columnwidth]{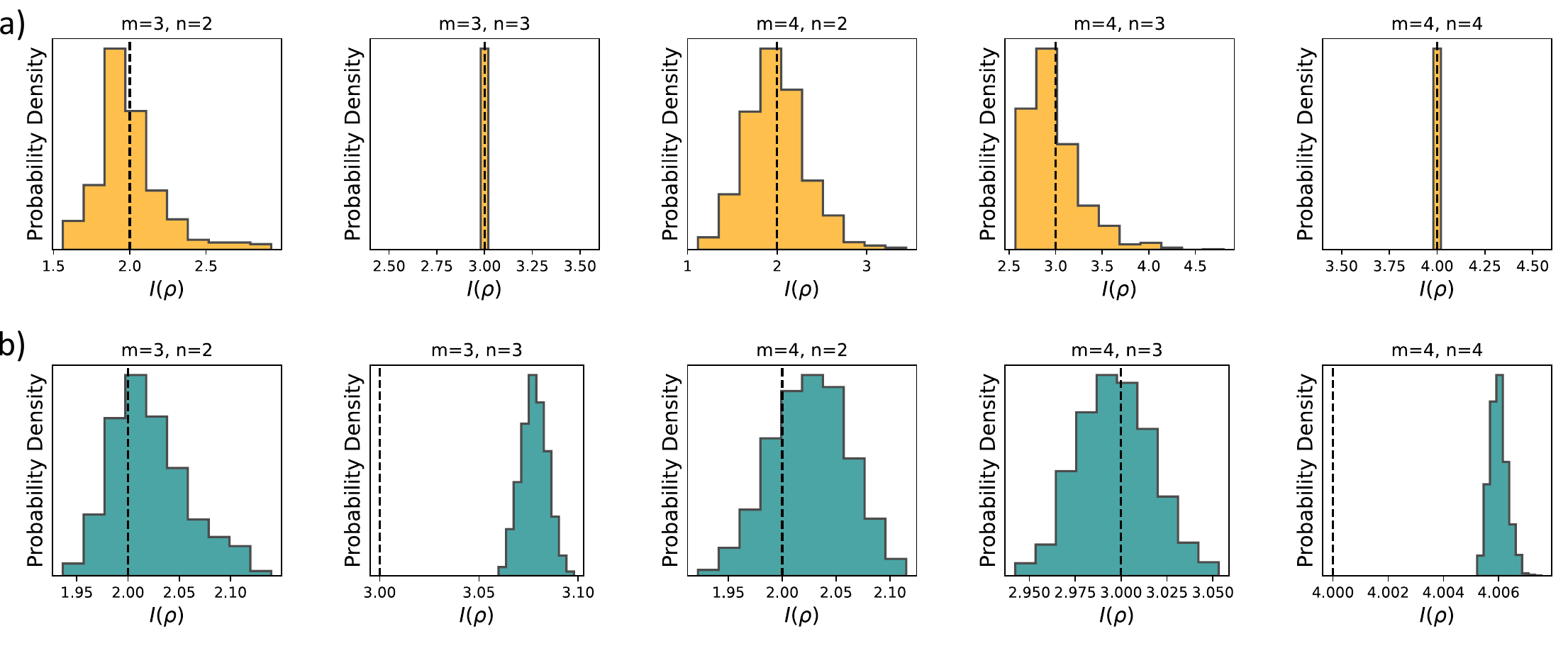}
    \caption{Histograms showing the deviations of the measured invariant \( I(\rho) \) from its ideal value under different noise models and \(\{m,n\}\) configurations. (a) Imperfections in the implementation of the observables due to mismatched beam splitter reflectivities (\(R = 0.52\)).
(b) Calibration errors in the detection efficiencies, modeled with a Gaussian distribution of standard deviation \(\delta = 0.06\). Each panel includes results for different $\{m,n\}$ configurations, with \( m = 3, 4 \) and \( n = 2, 3, 4 \).}

    \label{fig: Noise}
\end{figure}

In this section, we investigate how different kinds of imperfections influence the measurement of the Lie invariant quantity \( I(\rho) \) defined in the main text. In particular, we consider four different sources of noise: partial distinguishability of photons, multi-photon contributions from the source, imperfections in the implementation of the measurement of the observables \( O_i \), and calibration errors in pseudo-photon-number-resolving detectors. For each noise model, we simulate several levels of noise and average over 1000 Haar-random unitaries over different $\{m,n\}$ configurations, computing the output probabilities and mean photon numbers after the application of the observable operators required for the evaluation of \( I(\rho) \). The main results are the following. Regarding partial distinguishability, we see that it does not alter the measured values for $I(\rho)$, as we expect from the discussion of the previous section. As for the noise introduced by multi-photon contributions from the source, we see that its effect is just to shift the value of each measured invariant by a small positive quantity, which remains of order $\mathcal{O}(10^{-4})$ for values of $g^{(2)}(0)$ up to $g^{(2)}(0) =0.05$. We highlight that the effect of multi-photon contributions in the Lie invariants measurement can also depend on the detection system. Indeed, if post-selection in the output photon number is performed, so as to keep only those terms with a fixed number of detected photons, and losses in the apparatus are balanced, this small shift is not present. The main source of noise is then given by the other two effects. In Fig.~\ref{fig: Noise} we show histograms illustrating the deviations of the measured invariant values from their ideal counterparts across different \(\{m,n\}\) configurations. The two examples, shown in different colors, correspond to representative scenarios for each class of noise. The first case concerns imperfections in the implementation of the measurements of \( O_i \), where the interferometric scheme used to realize these measurements is affected by systematic deviations in the beam splitter reflectivities. In particular, all directional couplers in the Mach-Zehnder interferometers are assumed to have a fixed reflectivity \( R = 0.52 \), preventing ideal bar and cross transformations and introducing errors in the measurement process. The second case addresses detector-related imperfections in a pseudo-photon-number-resolving (PNR) detection system subject to calibration errors. Here, the quantum efficiencies associated with each output mode are misestimated due to imperfect calibration, introducing a mode-dependent distortion. This effect is modeled by sampling the relative detection efficiencies from a normal distribution with mean \( \eta_i = 1 \) and standard deviation \( \delta_i = 0.06 \), and applying the same realization across the entire ensemble of Haar-random unitaries.

\section{Measuring Lie Invariant Quantities for Inputs with More Than One Photon in a Mode}

\begin{figure}[ht!]
    \centering
    \includegraphics[width = 0.85\columnwidth]{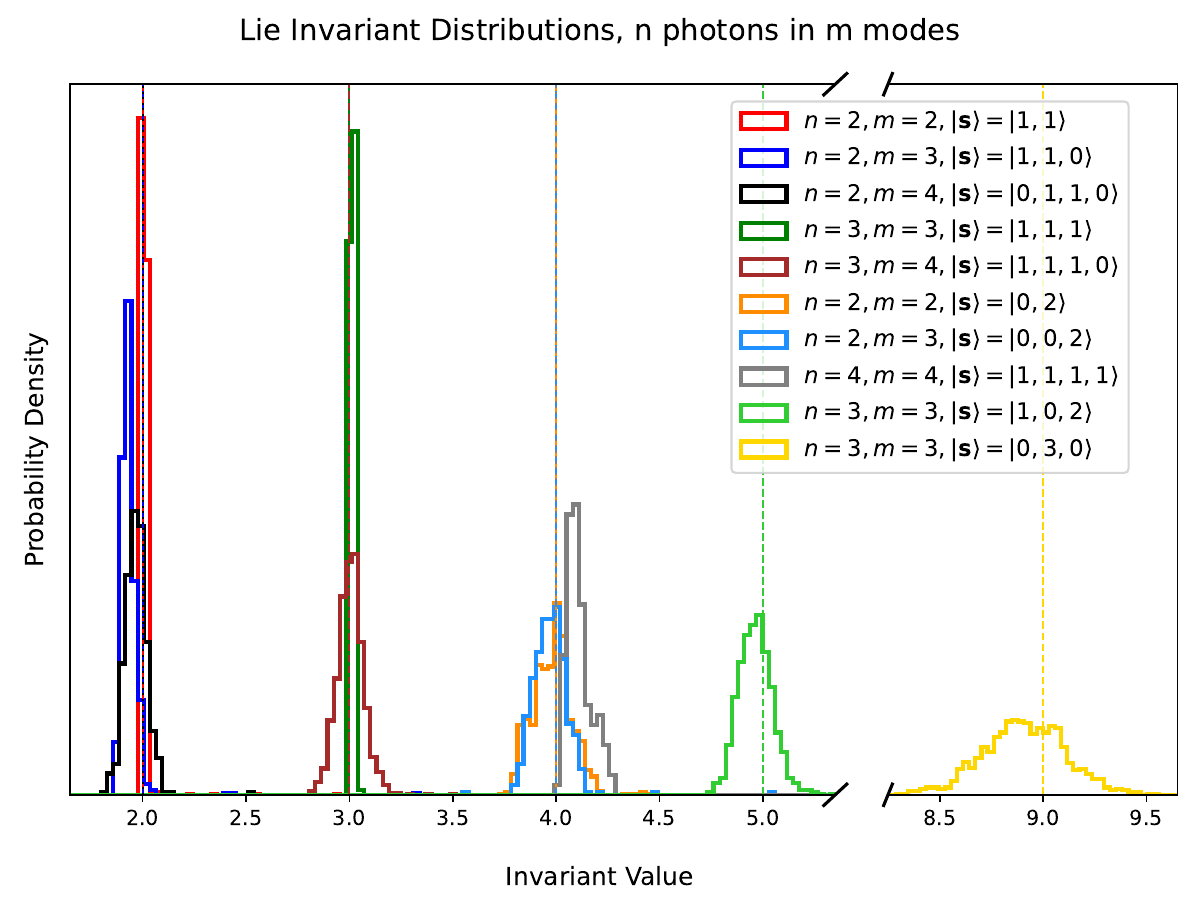}
    \caption{\textbf{Histograms of the measured values for the Lie invariant obtained with different configurations.} Here, we report the experimentally obtained distribution of the measured Lie invariants, for different configurations obtained by varying the number of photons $n$ and modes $m$, choosing a corresponding input Fock states $\ket{\bm{s}}$. Input Fock states featuring more than one photon per mode are obtained upon suitable post-selection, preceding the interferometer $U$ with a suitable state-preparation setup. For each color-coded configuration, the vertical lines correspond to the theoretically expected value of the invariant for each configuration.}
    \label{fig: Distribution}
\end{figure}

The invariance of the quantity \( I(\rho) \), as defined in the main text, was also tested for input states containing more than one photon in a given mode, such as \( \ket{\bm{s}} = \ket{2,0} \) in the \( n=2 \), \( m=2 \) scenario.  To generate these input states, the linear optical evolution implementing a Haar-random unitary transformation \( U \in SU(m) \) was preceded by an input state preparation stage. This stage consisted of a sequence of balanced beam splitters, exploiting the Hong--Ou--Mandel effect to probabilistically prepare the desired input state. Due to the non-deterministic nature of this preparation process, the output probability distribution was reconstructed via post-selection, considering only detection events in which all photons were detected in the output modes.
Table~\ref{tab:stats} summarizes the obtained results, reporting the average measured value of \( I(\rho) \) and its associated standard deviation across all implemented unitary transformations for each input state across different \(\{n,m\}\) configuration, as well as the total number of implemented unitary transformations. Additionally, Fig.~\ref{fig: Distribution} presents the distributions of the experimentally obtained values of \( I(\rho) \) for each input state.

\begin{table*}[ht!]
\centering
\begin{tabular}{|c|c|c|c|}
\hline
\textbf(n,m) & $\ket{\bm{s}}$ & $I(\rho)$ & \# U \\
\hline
(2,2)  & $\ket{1,1}$ & 2.008(40) & 1000 \\
\hline
(2,3)  & $\ket{1,1,0}$  & 1.936(73) & 500 \\ 
\hline
(2,2)  & $\ket{0,2}$  & 3.982(95) & 1000  \\ 
\hline
(2,3)  & $\ket{0,0,2}$  & 3.972(90) & 500  \\ 
\hline
(3,3)  & $\ket{1,1,1}$  & 3.012(6) & 1000 \\ 
\hline
(3,3)  & $\ket{1,0,2}$ & 4.962(82) & 1000  \\ 
\hline
(3,3)  & $\ket{0,3,0}$  & 8.90(20)& 1000 \\ 
\hline
(2,4)  & $\ket{0,1,1,0}$  & 1.970(57) & 400 \\ 
\hline
(3,4)  & $\ket{1,1,1,0}$  & 3.007(65) & 1250 \\ 
\hline
(4,4)  & $\ket{1,1,1,1}$  & 4.111(58) & 120  \\ 
\hline
\end{tabular}
\caption{Mean value and standard deviation of the measured values for the Lie invariant \( I(\rho) \) obtained for different input states \( \ket{\bm{s}} \) with varying numbers of photons \( n \) and modes \( m \), including also cases where more than one photon is injected into the same mode, obtained upon suitable post-selection. The values are averaged over multiple Haar-random unitary transformations \( U \in SU(m) \). In the last column, we report the total number of unitary transformations sampled for each configuration.}

\label{tab:stats}
\end{table*}

\section{Analytical derivation of the spectrum of the Projection of the Density Matrix onto the Subalgebra of Linear Optical Hamiltonians}

Given the set of operators $O_i$ defined in Eq.\ \eqref{eq:observables}, constituting a basis of the linear optical Hamiltonians for the evolution of a quantum state in a $m$-mode linear network, it is possible to define the projection of the density matrix $\rho$ of any $m$-mode quantum state onto the subalgebra of linear optical Hamiltonians as follows:
\begin{equation}
    \rho_T := \sum_{i=1}^{m^2} \mathrm{tr}(O_i \rho) O_i.
\end{equation}
According to \cite{invariants} the spectrum of this operator is invariant under linear optical evolution. In the following we are going to establish a way to experimentally test this statement in the $m=2$ and $m=3$ scenarios.
\subsection{2-mode scenario}
\begin{figure}[ht!]
    \centering
    \includegraphics[width = 0.65\columnwidth]{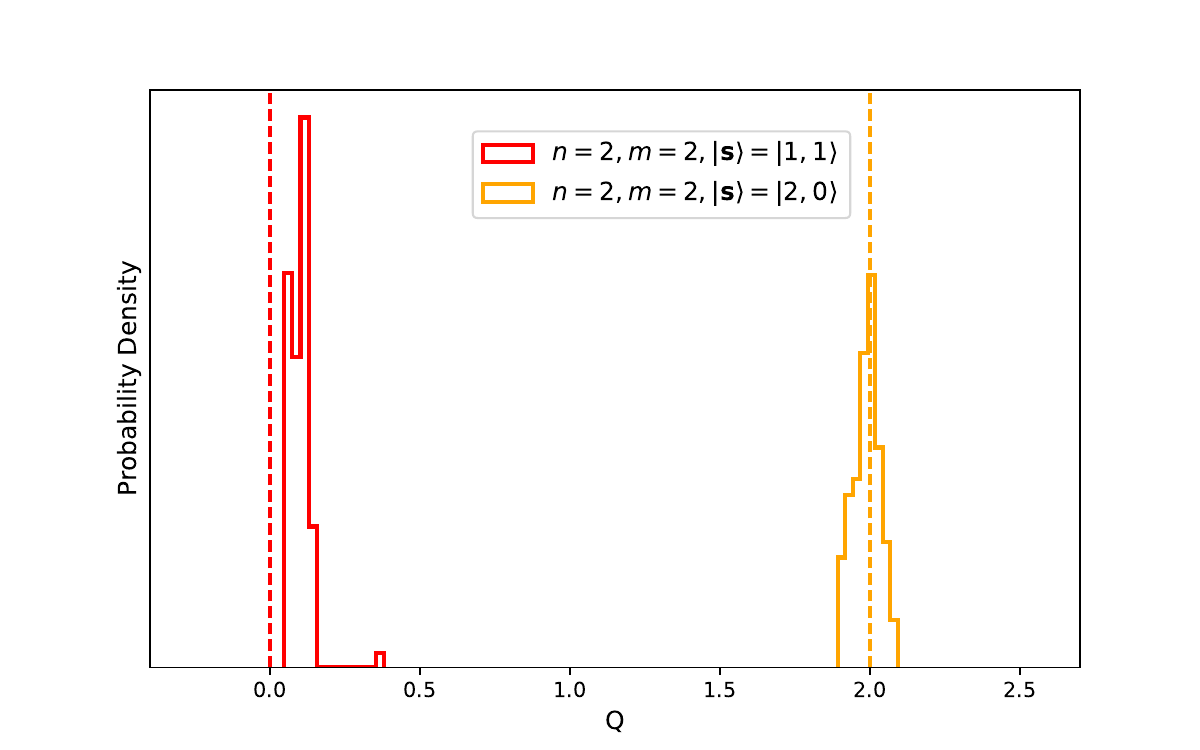}
    \caption{\textbf{Histograms of the measured values for quantity $Q$ as obtained with different configurations.} Here, we report the experimentally obtained distributions of the measured quantity $Q$ related to the spectrum of $\rho_T$, for the two different input Fock states $\ket{\bm{s}}$ with a fixed number of photons $n=2$ and modes $m=2$, across $1000$ different Haar-random unitary transformations $U \in SU(2)$. Input state $\ket{2,0}$ is obtained upon suitable post-selection, preceding the unitary transformation $U$ with a suitable state-preparation setup. For each color-coded configuration, the vertical lines indicate the theoretically expected value of the quantity $Q$.}
    \label{fig: eigenvalues_exp}
\end{figure}

To begin, we consider the problem of finding the eigenvalues of $\rho_T$ in the simplest case, where $\rho$ describes a $2$-mode state ($m=2$). In this case, we can rewrite $\rho_T$ explicitly as:
\begin{equation}\label{eq:rhoT_2m}
    \rho_T = \langle O_1^z\rangle_{\rho} n_1 + \langle O_2^z\rangle_{\rho} n_2 + \langle O_{12}^x\rangle_{\rho} \frac{1}{\sqrt{2}}(a_1^\dagger a_2 +a_2^\dagger a_1) + \langle O_{12}^y\rangle_{\rho}\frac{i}{\sqrt{2}}(a_1^\dagger a_2 -a_2^\dagger a_1) = N_1 n_1 + N_2 n_2 + R_{12} a_1^\dagger a_2 + R_{12}^* a_2^\dagger a_1,
\end{equation}
where we defined:
\begin{equation}
    \quad N_j = \langle O_j^z\rangle_{\rho} = \langle n_j \rangle_{\rho}, \quad R_{12} = \frac{1}{\sqrt{2}} (\langle O_{12}^x\rangle_{\rho}+i\langle O_{12}^y\rangle_{\rho}) = \frac{1}{2}(\langle (a_1^\dagger a_2 +a_2^\dagger a_1) \rangle_{\rho}- \langle (a_1^\dagger a_2 -a_2^\dagger a_1) \rangle_{\rho}) = \langle a_2^\dagger a_1 \rangle_{\rho}.
\end{equation}
The eigenstates of $\rho_T$ are of the form:
\begin{equation}\label{eq: states}
\ket{\psi} = \sum_{n_1,n_2} c_{n_1,n_2}\ket {n_1,n_2},
\end{equation}
and the action of the operators in Eq. \eqref{eq:rhoT_2m} on any basis state is given by:
\begin{equation}\label{eq: a_adagger}
\begin{aligned}
  & n_1 \ket {n_1,n_2} = n_1 \ket {n_1,n_2} \qquad  n_2 \ket {n_1,n_2} = n_2 \ket {n_1,n_2} \\
  & a_1^\dagger a_2 \ket {n_1,n_2} = \sqrt{(n_1+1)n_2}\ket{n_1+1,n_2-1} \qquad a_2^\dagger a_1 \ket {n_1,n_2} = \sqrt{(n_1(n_2+1)}\ket{n_1-1,n_2+1}.
\end{aligned}    
\end{equation}
Note that all the operators in $\rho_T$ conserve the total number of photons. It follows that $\rho_T$ is block-diagonal in Fock basis and the eigenvectors for each block are given by the following subset of the states in Eq.\ \eqref{eq: states}:
\begin{equation}\label{eq:eigenvector2m}
    \ket{\psi_n} = \sum_{k=0}^n c_k\ket{n-k,k},
\end{equation}
where we denote now with $n = n_1+n_2$ the total number of photons (corresponding to each block of $\rho_T$), and we have $\sum_k\abs{c_k}^2 = 1$. To find the eigenstates and the corresponding eigenvalues we write the eigenvalue equation:
\begin{equation}\label{eq: eigenvalue}
    \rho_T \ket{\psi_n} = \lambda_n \ket{\psi_n},
\end{equation}
which corresponds to (Eqs.\ \eqref{eq:rhoT_2m} and \eqref{eq:eigenvector2m}):
\begin{equation}
    (N_1 n_1 + N_2 n_2 + R_{12} a_1^\dagger a_2 + R_{12}^* a_2^\dagger a_1) \sum_{k=0}^n c_k\ket{n-k,k} = \lambda_n \sum_{k=0}^n c_k\ket{n-k,k}.
\end{equation}
Applying Eq.\ \eqref{eq: a_adagger} we get the following system of equations for the coefficients $\{c_k\}$ and the eigenvalue $\lambda_n$:
\begin{equation}
\begin{cases}    
&(N_1(n-k)+N_2k)c_k+R_{12}\sqrt{(n-k)(k+1)}c_{k+1}+R_{12}^*\sqrt{k(n-k+1)}c_{k-1} = \lambda_n c_k \qquad \forall k, \\
& \sum_k\abs{c_k}^2 = 1.
\end{cases}
\end{equation}
It can be shown that the system has a solution for a set of $n+1$ eigenvalues $\{ \lambda_n^j \}$:
\begin{equation}
\lambda_n^j=\frac{n}{2}(N_1+N_2)+j\sqrt{(N_1-N_2)^2+4\abs{R_{12}}^2} \qquad j \in \{-\frac{n}{2}, -\frac{n}{2} + 1, \dots, \frac{n}{2}-1, \frac{n}{2}\}.
\end{equation}
We now want to test if the spectrum of $\rho_T$ is invariant under linear optical evolution. In order to do so we consider as possible input states the Fock states of $2$ photons in $2$ modes: $\ket{\bm{s}} = \ket{1,1}$ and $\ket{\bm{s}} = \ket{2,0}$. We know that the total number of photons $N_1 + N_2$ must be conserved after the evolution, and this requirement is indeed imposed in the experimental procedure as a normalization condition. For this reason, the non-trivial part in the spectrum is actually given by the quantity:
\begin{equation}
    Q = \sqrt{(N_1-N_2)^2+4\abs{R_{12}}^2}.
\end{equation}
Recalling that $N_1 = \langle n_1 \rangle_{\rho}$, $N_2 = \langle n_2 \rangle_{\rho}$ and $R_{12} = \langle a_2^\dagger a_1  \rangle_{\rho}$ and that for Fock states we have $\langle{n_j}\rangle_{\ket{\bm{s}}} = n_j$ and $ \langle a^\dagger_j a_k \rangle_{\ket{\bm{s}}} = 0$, we obtain
\begin{equation}\label{eq:Q}
\begin{aligned}
  &\ket{\bm{s}} = \ket{1,1} \implies N_1 = 1, N_2 = 1, R_{12}=0 \implies Q_{\text{theo}} = 0, \\
  &\ket{\bm{s}} = \ket{2,0} \implies N_1 = 2, N_2 = 0, R_{12}=0 \implies Q_{\text{theo}} = 2. 
\end{aligned}
\end{equation}
To demonstrate that the spectrum of $\rho_T$ is indeed invariant, we experimentally reconstruct the quantity $Q$ on the set of evolved states by measuring $N_1$, $N_2$, and $R_{12}$. These quantities can be directly obtained from experimental data, as they are linear combinations of the expectation values of the Lie observables $\expval{O_i}_{\rho}$. We report in Tab. \ref{tab:eig} the measured value of \( Q \) and its associated standard deviation across all the unitary transformations implemented for the two input states. In Fig. \ref{fig: eigenvalues_exp} we show histograms for the measured values of $Q$ for the two inputs, comparing them with the expected theoretical value of Eq.\ \eqref{eq:Q}. 
Interestingly, we can establish a direct link between the quantity \( Q \) and the single-particle coherency matrix $\mathbf{\Gamma}^{(1)}$. In the 2-mode case, $\mathbf{\Gamma}^{(1)}$ is a \( 2 \times 2 \) Hermitian matrix with elements \( \Gamma^{(1)}_{jk} = \langle a_j^\dagger a_k \rangle_{\rho} \):
\begin{equation}
\mathbf{\Gamma}^{(1)} = \begin{pmatrix}
\langle a_1^\dagger a_1 \rangle_{\rho} & \langle a_1^\dagger a_2 \rangle_{\rho} \\
\langle a_2^\dagger a_1 \rangle_{\rho} & \langle a_2^\dagger a_2 \rangle_{\rho}
\end{pmatrix}.
\end{equation}
Defining \( N_1 = \langle a_1^\dagger a_1 \rangle_{\rho} \), \( N_2 = \langle a_2^\dagger a_2 \rangle_{\rho} \), and \( R_{12} = \langle a_2^\dagger a_1 \rangle_{\rho} \), we obtain:
\begin{equation}
\operatorname{Tr}[\mathbf{\Gamma}^{(1)}] = N_1 + N_2, \qquad 
\operatorname{Tr}[\mathbf{\Gamma}^{(1)\dagger}\mathbf{\Gamma}^{(1)}] = N_1^2 + N_2^2 + 2|R_{12}|^2.
\end{equation}
Now consider the quantity \( Q^2 \):
\begin{equation}
Q^2 = (N_1 - N_2)^2 + 4|R_{12}|^2= N_1^2 + N_2^2 - 2N_1N_2 + 4|R_{12}|^2= 2\operatorname{Tr}[\mathbf{\Gamma}^{(1)\dagger} \mathbf{\Gamma}^{(1)}] - \left( \operatorname{Tr}[\mathbf{\Gamma}^{(1)}] \right)^2.
\end{equation}
This can be written compactly as:
\begin{equation}
Q^2 = 2\left\| \mathbf{\Gamma}^{(1)} - \tfrac{1}{2} \operatorname{Tr}[\mathbf{\Gamma}^{(1)}] \mathbb{I} \right\|^2 = 2\| \mathbf{\Gamma^{(1)}_{\mathrm{tl}}} \|^2,
\end{equation}
where \( \mathbf{\Gamma^{(1)}_{\mathrm{tl}}} \) is the traceless part of the matrix, defined as \( \mathbf{\Gamma^{(1)}_{\mathrm{tl}}} = \mathbf{\Gamma}^{(1)} - \tfrac{1}{2} \operatorname{Tr}[\mathbf{\Gamma}^{(1)}] \mathbb{I} \). Thus, \( Q \) quantifies the norm of the traceless component of the coherency matrix and captures the non-trivial structure of coherences and populations beyond the conserved total photon number.

As expected, in the 2-mode case there are exactly two independent invariants under linear optics (based on first-order correlators $\langle a_j^\dagger a_k \rangle$): the total photon number \( N_1 + N_2 \) and the quantity \( Q \). Alternatively, one can express these invariants through the eigenvalues \( \lambda_{\pm} \) of $\mathbf{\Gamma}^{(1)}$, which actually coincide with those of the first diagonal block of \( \rho_T \):
\begin{equation}
\lambda_{\pm} = \frac{1}{2}(N_1 + N_2) \pm \frac{Q}{2}.
\end{equation}
This shows that describing the state using \( \{Q, N_1 + N_2\} \) is equivalent to using \( \{\lambda_+, \lambda_-\} \): either way, one has exactly two independent invariants, matching the number of modes.

Finally, recalling that the invariant \( I(\rho) = \operatorname{Tr}[\mathbf{\Gamma}^{(1)\dagger} \mathbf{\Gamma}^{(1)}] \), we find:
\begin{equation}
Q^2 = 2I(\rho) - (N_1 + N_2)^2,
\end{equation}
which confirms that any pair of invariants constructed from either \( \rho_T \) or $\mathbf{\Gamma}^{(1)}$ is ultimately equivalent to the pair \( \{ I(\rho), N_1 + N_2 \} \).

\subsection{3-mode scenario}
\begin{figure}[ht!]
    \centering
    \includegraphics[width = 0.999\columnwidth]{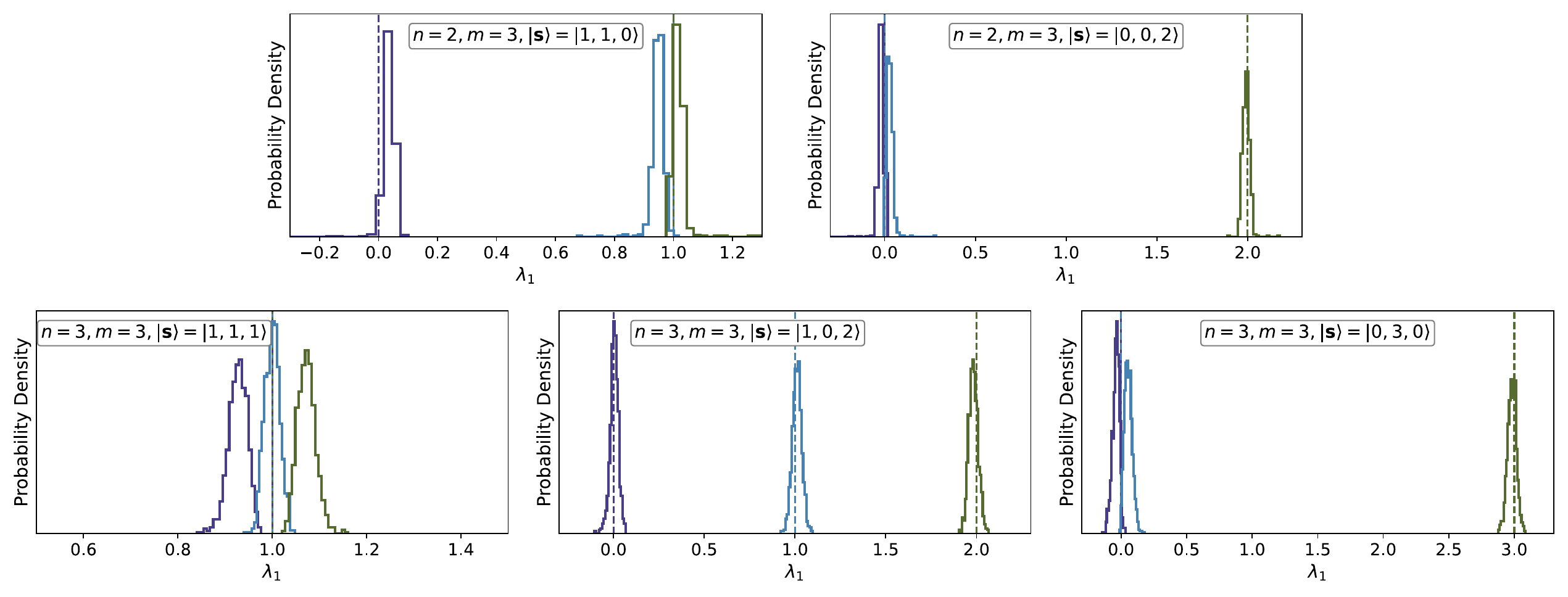}
    \caption{\textbf{Histograms of the measured values for the eigenvalues $\{\lambda_1^j\}$ obtained from experimental data for different three-mode configurations.} Here, we report the distribution of the experimentally measured eigenvalue $\lambda_1$ corresponding to the spectrum of $\rho_T$ reduced to the first diagonal block ($n=1$) across $\mathcal{O}(10^3)$ different Haar-Random unitary transformations. These eigenvalues correspond to those of the coherency matrix $\mathbf{\Gamma}^{(1)}$. Different input Fock states $\ket{\bm{s}}$ are considered, with $n=2$ and $n=3$ photons in $m=3$ modes. Input states featuring more than one photon per mode are obtained via suitable post-selection, preceding the unitary transformation $U$ with an appropriate state-preparation setup. For each color-coded configuration, vertical lines indicate the expected theoretical values of the eigenvalues $\{\lambda_1^j\}$.}
    \label{fig: lambda1_3m}
\end{figure}

As an alternative scenario, we now consider the case where $\rho$ describes a three-mode state ($m = 3$). In this case, we define:
\begin{equation}\label{eq:rhoT_3m}
\begin{aligned}
        \rho_T := & \langle O_1^z\rangle_{\rho} n_1 + \langle O_2^z\rangle_{\rho} n_2 + \langle O_3^z\rangle_{\rho} n_3 + \langle O_{12}^x\rangle_{\rho} \frac{1}{\sqrt{2}}(a_1^\dagger a_2 +a_2^\dagger a_1) + \langle O_{13}^x\rangle_{\rho}\frac{1}{\sqrt{2}}(a_1^\dagger a_3 +a_3^\dagger a_1) \\ 
        &\quad+ \langle O_{23}^x\rangle_{\rho}\frac{1}{\sqrt{2}}(a_2^\dagger a_3 +a_3^\dagger a_2) + \langle O_{12}^y\rangle_{\rho} \frac{i}{\sqrt{2}}(a_1^\dagger a_2 -a_2^\dagger a_1) + \langle O_{13}^y\rangle_{\rho}\frac{i}{\sqrt{2}}(a_1^\dagger a_3 -a_3^\dagger a_1) \\ 
        &\quad+ \langle O_{23}^y\rangle_{\rho}\frac{i}{\sqrt{2}}(a_2^\dagger a_3 -a_3^\dagger a_2) \\
        =& N_1 n_1 + N_2 n_2 + N_3 n_3 + R_{12} a_1^\dagger a_2 + R_{12}^* a_2^\dagger a_1 + R_{13} a_1^\dagger a_3 + R_{13}^* a_3^\dagger a_1 \\ 
        &\quad+ R_{23} a_2^\dagger a_3 + R_{23}^* a_3^\dagger a_2,
\end{aligned}
\end{equation}
where again we defined:
\begin{equation}
 N_j = \langle O^z_j \rangle_{\rho}=\langle n_j \rangle_{\rho}, \qquad R_{jk} = \frac{1}{\sqrt{2}} (\langle O_{jk}^x\rangle_{\rho}+i\langle O_{jk}^y\rangle_{\rho}) = \langle a_k^\dagger a_j\rangle_{\rho}.     
\end{equation}
Once again, since the operators $O_i$ preserve the total number of photons, the eigenstates of $\rho_T$ are superpositions of three-mode Fock states with a fixed photon number $n$:
\begin{equation}
    \ket{\psi_n} = \sum_{l=0}^n \sum_{k=0}^l  c_{l,k}\ket{n-l,l-k,k},
\end{equation}
with $\sum_{l,k}\abs{c_{l,k}}^2 = 1$. In the three-mode case, we restrict our analysis to the eigenvectors and eigenvalues corresponding to $n=1$. This gives for the eigenstates:
\begin{equation}
    \ket{\psi_1} = c_{0,0}\ket{1,0,0}+ c_{1,0} \ket{0,1,0}+ c_{0,1} \ket{0,0,1}.
\end{equation}
In this case we limit our analysis to the first diagonal block of $\rho_T$. In order to do so we impose the eigenvalue equation (Eq.\ \eqref{eq: eigenvalue}) for $n=1$. By using expressions analogous to the ones in Eq.\ \eqref{eq: a_adagger} for the action of $n_j$ and $a^\dagger_j a_k$ operators appearing in $\rho_T$ (Eq.\ \eqref{eq:rhoT_3m}), we obtain the following system of equations for the coefficients $\{c_{l,k}\}$ and the eigenvalue $\lambda_1$:
\begin{equation}\label{system 3m lambda 1}
\begin{cases}
N_1 c_{0,0} + R_{12}  c_{1,0} +  R_{13}  c_{0,1} = \lambda_1 c_{0,0}, \\
R^*_{12} c_{0,0} + N_2  c_{1,0} +  R_{23}  c_{0,1} = \lambda_1 c_{1,0}, \\
R^*_{13} c_{0,0} + R^*_{23}  c_{1,0} +  N_3  c_{0,1} = \lambda_1 c_{0,1}.
\end{cases}
\end{equation}
For any input three-mode Fock state $\ket{\bm{s}}$, we always have $R_{jk} = 0$, making the system diagonal and yielding the eigenvalues:
\begin{equation}\label{eq:lambda1}
    \lambda_1^j = N_j = \langle n_j \rangle_{\ket{\bm{s}}} = n_j , \qquad j = 1,2,3,
\end{equation}
where $\{n_j\}$ denote the input state occupation numbers.
We now investigate whether these eigenvalues remain invariant under linear optical evolution. To do so, we compute, for each output state, the matrix $M$:
\begin{equation}\label{eq:matrix 3m lambda1}
M = \begin{bmatrix}
N_1 & R_{12} & R_{13} \\
R^*_{12} & N_2 & R_{23} \\
R^*_{13} & R^*_{23} & N_3
\end{bmatrix},
\end{equation}
where $N_j$ and $R_{jk}$ are now determined from experimental data for the different configurations. The eigenvalues $\{\lambda_1^j\}$ are then obtained by numerically diagonalizing the matrix $M$. We report in Tab. \ref{tab:eig} the measured eigenvalues $\{\lambda_1^j\}$ and the associated standard deviation across all the unitary transformations implemented for the different input states for the $m=3$ scenario, with $n=2$ or $n=3$, including also states with more than a photon per mode. Figure \ref{fig: lambda1_3m} presents histograms of the experimentally obtained eigenvalues for the different configurations. In the figure the results are compared with the expected theoretical values from Eq.\ \eqref{eq:lambda1}. 

Once again we can link these results to the coherency matrix $\mathbf \Gamma^{(1)}$. In this case we simply note that, as in the $m=2$ case, the eigenvalues of $\mathbf \Gamma^{(1)}$ are exactly $\{\lambda_1^j\}$, since $M$ is exactly the complex conjugate of $\mathbf \Gamma^{(1)}$. Moreover, since we know that one can get at most $m$ independent invariant quantities from first order correlators $\langle a_k^\dagger a_j\rangle$, all the information related to invariant quanities provided by $\rho_T$ in this scenario is expected to be known if one consider only its first block, in an equivalent way to the $m=2$ case shown in the previous section.

\begin{table*}[ht!]
\centering
\begin{tabular}{|c|c|c|c|}
\hline
\textbf(n,m) & $\ket{\bm{s}}$ & Q / $\{\lambda_1^j\}$ & \# U \\
\hline
(2,2)  & $\ket{1,1}$ & 0.100(38) & 1000 \\
\hline
(2,2)  & $\ket{0,2}$  & 1.988(45) & 1000  \\ 
\hline
\hline
(2,3)  & $\ket{1,1,0}$  & $\mqty(0.035(29) \\ 0.947(24) \\ 1.018(32))$ & 500 \\
\hline
(2,3)  & $\ket{0,0,2}$  & $\mqty(-0.021(25) \\ 0.028(20) \\ 1.992(21))$ & 500 \\
\hline
(3,3)  & $\ket{1,1,1}$ & $\mqty(0.927(20)\\ 1.000(16)\\ 1.073(19))$ & 1000\\ 
\hline
(3,3)  & $\ket{1,0,2}$ & $\mqty(0.004(23)\\ 1.011(26)\\ 1.984(26))$ & 1000\\ 
\hline
(3,3)  & $\ket{0,3,0}$ & $\mqty(-0.038(27)\\ 0.056(29)\\  2.98233)$ & 1000\\
\hline
\end{tabular}
\caption{Mean value and standard deviation of the measured measured quantities related to the spectrum of $\rho_T$, which are $Q$ for $m=2$ and $\{\lambda_1^j\}$ for $m=3$. In the table we include also cases where more than one photon is injected into the same mode, obtained upon suitable post-selection. The values are averaged over multiple Haar-random unitary transformations \( U \in SU(m) \). In the last column, we report the total number of unitary transformations sampled for each configuration.}

\label{tab:eig}
\end{table*}

\bibliography{biblio_arxiv.bib}
\bibliographystyle{apsrev4-1}